\documentclass{plantillaTFM}
\usepackage{fvextra}
\usepackage[utf8]{inputenc}
\usepackage{newunicodechar}
\newunicodechar{⁻}{$^{-}$}



\begin{document}
\begin{titlepage} 
\begin{center} 

\includegraphics[width=10cm]{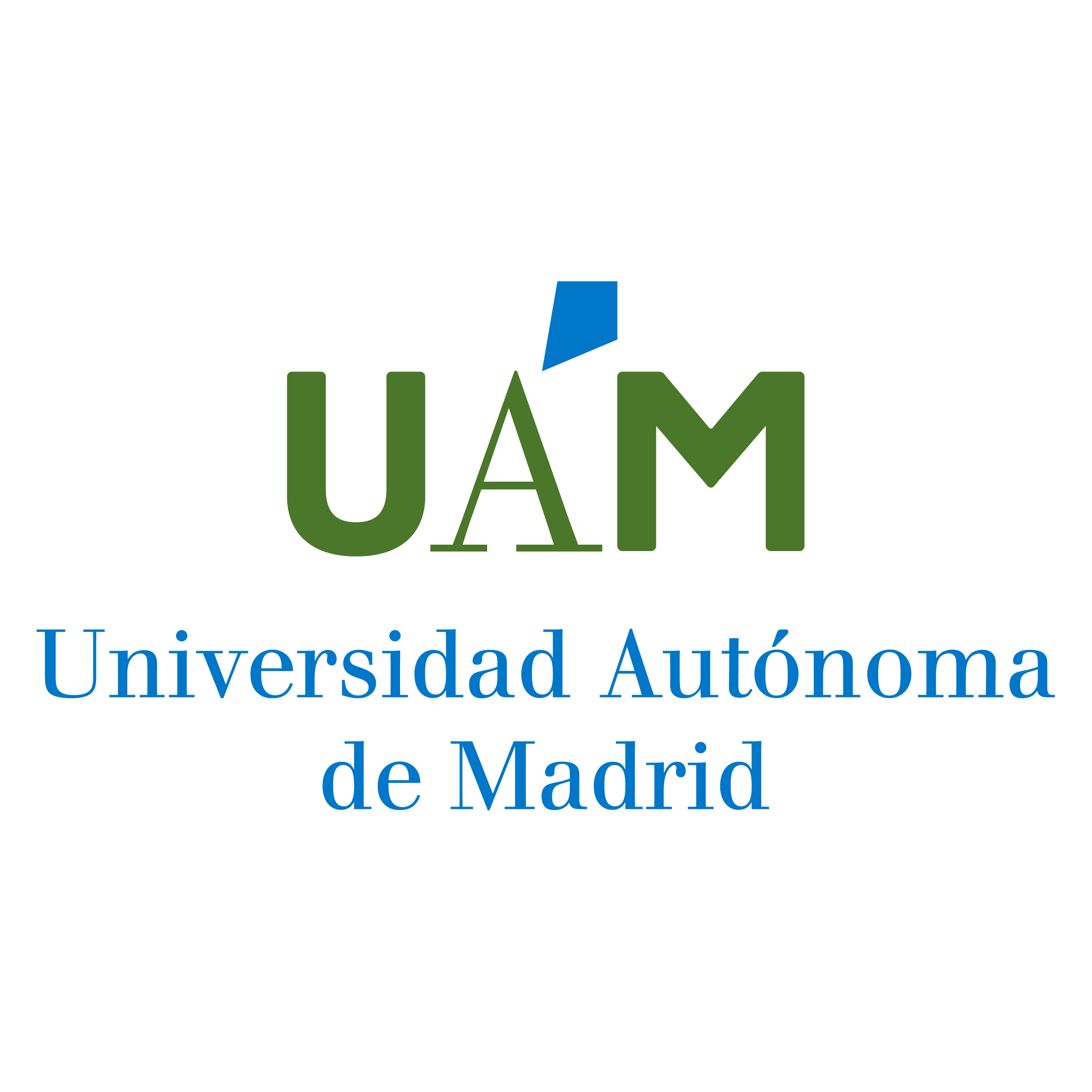}\\ 
\vspace*{3\baselineskip}
{TRABAJO FIN DE MÁSTER \\ [0.2cm] Master in theoretical physics - Speciality Astrophysics\\ Academic Year: 2024/2025 \\[0.8cm]}
{\Large\textbf{MOLECULAR CLOUDS AS A TOOL TO PLACE CONSTRAINS IN SUB-GEV DARK MATTER AND PRIMORDIAL BLACK HOLES}} \\[1.8cm]
 Asier Salces Pérez\\[0.3cm]
Supervisors: Dr. Pedro de la Torre Luque (UAM \& IFT-UAM-CSIC) \\ Dr. Miguel Ángel Sánchez Conde (UAM \& IFT-UAM-CSIC)\\[0.3cm]

Place:  Department of Theoretical Physics, Faculty of Science at Instituto de Física Teórica (IFT)
\end{center}
\end{titlepage}

\thispagestyle{empty}
\vspace*{0cm}
\begin{verse}
\itshape
Do not go gentle into that good night, \\
Old age should burn and rave at close of day; \\
Rage, rage against the dying of the light. \\[1em]

Though wise men at their end know dark is right, \\
Because their words had forked no lightning they \\
Do not go gentle into that good night.
\end{verse}

\textit{— Dylan Thomas (1914–1953)}
\clearpage

\thispagestyle{empty}
\section*{\Huge Acknowledgments}

\vspace*{9cm}
\begin{center}
\itshape
To my beloved ones,\\
whose light has guided me in every step.\\
To my cousin, always with us and forever\\ remembered. \\[1.2em]

To my supervisors, Pedro and Miguel \'Angel,\\
for their wisdom and patience,\\
and to the DAMASCO group,\\
a source of support and inspiration.
\par
\end{center}

\clearpage


\tableofcontents
\thispagestyle{empty} 
\newpage
\setcounter{page}{1} 
\newcommand{\bra}[1]{\langle#1|} 
\newcommand{\ket}[1]{|#1\rangle} 
\newcommand{\braket}[2]{ \langle #1 | #2 \rangle} 


\addcontentsline{toc}{section}{Abstract}
\section*{\centering Abstract}
Despite decades of direct and indirect searches for Dark Matter (DM) within the Weakly Interacting Massive Particle (WIMP) framework there is still no conclusive results in such mass range (GeV-TeV). This has motivated the exploration of alternatives beyond the traditional window, including both new particle candidates and macroscopic objects. Two particularly well motivated scenarios are sub-GeV DM and Primordial Black Holes (PBHs). Molecular clouds (MCs), typically studied as sites of star formation, can also serve as astrophysical laboratories to probe these candidates through their ionization rates. Observations show ionization levels exceeding expectations from known CR fluxes, pointing to an additional ionizing component. Here, we consider electrons and positrons from annihilating and decaying MeV DM particles, as well as Hawking radiation from evaporating PBHs, as possible contributors. A theoretical framework is implemented to model the transport driven by energy losses of these particles within the clouds. By comparing predicted ionization rates with observations, conservative constraints are set on the thermally averaged cross section $\langle\sigma v\rangle$, decay lifetime $\tau$ and PBH abundances $f_{PBH}$. The analysis assumes all the observed ionization comes from DM and adopts a 95\% confidence level. Results show that, even in the most conservative case of local MCs such as L1551, these constraints are very close to the most competitive bounds from X-ray observations, while inner-Galaxy clouds like DRAGON or G1.4--1.8+87 provide stronger limits, sometimes improving X-ray and cosmological constraints. For sub-GeV DM, MCs exclude parameter space competitive with the one tested by NuSTAR, INTEGRAL, or Voyager, especially below $\sim$30 MeV. In the PBH case, asteroid-mass black holes are restricted to a low fraction of DM, with optimistic scenarios getting close to the strongest limits. This work demonstrates the potential of MCs as a novel observable in indirect DM searches
\newpage
\addcontentsline{toc}{section}{Resumen}
\section*{\centering Resumen}
A pesar de décadas de esfuerzos en búsquedas directas e indirectas de materia oscura en el rango de las Partículas Masivas Débilmente Interactivas (WIMPs, GeV--TeV), los resultados siguen siendo inconclusos. Esto ha motivado la exploración de posibilidades alternativas, tanto en forma de nuevos candidatos de partículas como de objetos macroscópicos. En particular, dos escenarios prometedores son la materia oscura sub-GeV y los agujeros negros primordiales (PBHs). Con el fin de imponer restricciones sobre sus propiedades, se propone un nuevo observable: las nubes moleculares. Estos objetos, habitualmente estudiados como regiones de formación estelar, pueden también servir como laboratorios astrofísicos para explorar posibles candidatos de materia oscura a través de sus tasas de ionización. El uso de nubes moleculares viene motivado por las inesperadamente altas tasas de ionización observadas, que superan las predichas por las poblaciones estándar de rayos cósmicos, apuntando a un componente ionizante adicional. En este trabajo se consideran electrones y positrones provenientes de la aniquilación y decaimiento de partículas de materia oscura en el rango MeV, así como la radiación de Hawking de PBHs evaporándose. Se implementa un marco teórico para modelar el transporte dominado por pérdidas de energía de estas partículas dentro de las nubes. Comparando las tasas predichas con las observadas, se obtienen límites conservadores sobre $\langle\sigma v\rangle$, el tiempo de vida $\tau$ y la abundancia $f_{PBH}$, asumiendo de forma conservadora que toda la ionización procede de la materia oscura y adoptando un 95\% de confianza. Los resultados muestran que, incluso en el caso más conservador de nubes moleculares locales como L1551, estas restricciones están muy cerca de los límites más competitivos obtenidos a partir de observaciones de rayos X, mientras que nubes más internas como DRAGON o G1.4--1.8+87 ofrecen límites más estrictos, en algunos casos superando observaciones de rayos X y pruebas cosmológicas. Para la materia oscura sub-GeV, las nubes moleculares excluyen regiones del espacio de parámetros competitivas con NuSTAR, INTEGRAL o Voyager, especialmente por debajo de $\sim$30 MeV. En el caso de los PBHs, los de masa asteroidal quedan limitados a una fracción muy baja, acercándose en el escenario optimista a los límites más estrictos. De esta forma, se resalta el potencial de las nubes moleculares para imponer restricciones en búsquedas indirectas de materia oscura.

\newpage

\section{Introduction: Dark Matter}
From astrophysical and cosmological observations \cite{bertone2018history} it has been noticed that the total amount of mass required to explain certain processes in the universe, such as the rotation curves of galaxies, the dynamics of galaxy clusters, and the anisotropies of the cosmic microwave background, is much greater than what can be accounted for by the luminous matter it is detected directly. This indicates the presence of additional mass that cannot be easily observed. From what is known, such mass must not absorb or emit light (or, maybe if it is composed of particles, its coupling with photons should be exceptionally small) since it cannot be detected through its interaction with electromagnetic radiation. The only known interaction is through the gravitational force. In fact, all the current evidence for its existence comes from gravitational outcomes; dynamical effects on galaxies and galaxy clusters, lensing of light and even the primordial density perturbations in the early universe.
\subsection{Reasons to believe in its existence}
\subsubsection{Galactic rotational curves}
The circular velocity profiles of the objects within a galaxy are a great way to infer the total mass distribution of a galaxy. From the earliest spectroscopic measurements of Andromeda's rotation at the beginning of the XX century (Wolf, Slipher) \cite{wolf1914vierteljahresschr, slipher1914detection}, Francis Pease found that orbital speeds stayed almost constant far from the center implying that the observed mass was not enough to account for the measurement. In the following 20 years, studies from  Knut Lundmark \cite{1930MeLuF.125....1L} and Horace Babcock \cite{babcock1939rotation} found substantially high mass-to-light rations in several galaxies at large radii and reinforced their beliefs that large amounts of "dark matter" like comets, dead stars, dark MCs and so on should be present.  Babcock even had some others considerations \cite{babcock1939rotation}:\\

\textit{"[...] absorption plays a very important role in the outer portion
 of the spiral, or, perhaps, that new dynamical considerations are required, which
 will permit of a smaller mass in the outer parts."}\\

 After the World War II, 21 cm radio observations reinforced the flat-curve phenomenon \cite{van1957rotation} and a great study from  Franz Kahn and Lodewijk Woltjer \cite{kahn1959intergalactic} lead to the conclusion that the local group hinted an enormous amount of unseen mass. It was finally in 1970, when these accumulating discrepancies were at their peak, when Vera Rubin and Kent Ford published the famous paper at which was measured the rotational curve of M31 much more detailed \cite{rubin1970rotation}. Several authors showed that these curves could not be explained by modifying the weights of the observed mass and a new component should be surrounding the galaxies \cite{freeman1970disks}.
\subsubsection{Galaxy clusters}
Back in 1933, Fritz Zwicky, assuming that the Coma Cluster is an isolated system, applied the virial theorem relating the average velocity of the objects with its gravitational potential in order to subtract the total mass of the system \cite{zwicky1933rotverschiebung}. Since this method does not rely on luminous mass, he obtained a much higher value than with previous methods, about $\gtrsim 500$ mass-to-light ratio. This lead him to conclude that a vast amount of “dark matter” should be present \cite{zwicky1933rotverschiebung}:\\

\textit{"If this would be confirmed, we would get the surprising result that dark matter
is present in much greater amount than luminous matter."}\\

Works from Sinclair Smith reinforced this result and the attempts of Hubble and Holmberg to reevaluate the observations were not satisfactory \cite{hubble1982realm, holmberg1940clustering}. Finally, Wolf suggested that the gas could be ionized, but X-ray observations \cite{meekins1971physical} ruled out the possibility. The amount of gas detected was insufficient to provide enough gravitational binding to keep the Coma cluster stable. After this possibility was discarded, more exotic possibilities were proposed (massive collapsed objects, HI snow-balls, etc.). However, measurements of primordial abundances of elements \cite{burles2001big, hinshaw2013nine}, 
made clear that baryonic matter content of the universe was only about 20\%, clearly insufficient to account for the observed gravitational effects. DM as a form of non-baryonic matter was becoming the strongest theory that could explain all the discrepancies.
\subsubsection{Cosmological observations}
\begin{figure}[H]
    \centering
    \includegraphics[width=0.5\linewidth]{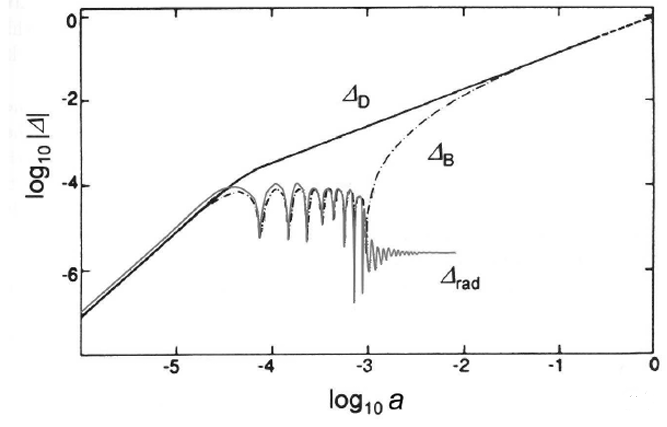}
    \caption{The evolution of the density fluctuations of baryonic matter ($\Delta_B$), radiation ($\Delta_\gamma$), and DM ($\Delta_D$) in a cold DM scenario, as a function of the scale factor $a$ (normalized so that $a = 1$ today). Here, $\Delta$ denotes the relative density contrast $\Delta \equiv \delta\rho / \rho$, which measures the fractional deviation of the local density from its average value. It illustrates that DM fluctuations grow much earlier and more efficiently than baryonic ones, since baryons were tightly coupled to photons before recombination and thus could not collapse into structures until the universe cooled enough for neutral atoms to form.}\cite{coles2003cosmology}
    \label{cmbfluc}
\end{figure}

One of the strongest arguments in favor of the existence of DM is the footprint left in the Cosmic Microwave Background (CMB) \cite{2020}. An inspection of the CMB density perturbations (see Figure \ref{cmbfluc}) shows that baryonic fluctuations alone cannot account for the observed large-scale structure of the Universe. Because baryons interact with photons, their density perturbations could not grow significantly before recombination. DM, however, being non-interacting with radiation, could decouple from the plasma well before recombination, allowing its perturbations to grow earlier. These fluctuations acted as gravitational wells, into which baryons later fell after recombination. 
Such DM could consist of PBHs formed during the inflationary epoch \cite{garcia1996density}, or of a new type of particle that does not interact electromagnetically and therefore decouples before recombination. This early growth of DM perturbations promoted baryonic density fluctuations to the level required for the subsequent formation of galaxies, clusters, and the cosmic web observed today.

The most recent observations from Planck \cite{2020} have confirmed that the universe is flat $k \simeq 0$ and it is dominated by DM and dark energy. In particular, the abundance of DM, expressed in terms of the cosmological density parameters is
\begin{equation}
    \Omega_{CDM}h^2 = \frac{\rho_{DM}}{\rho_c} = 0.1196 \pm 0.0031
\end{equation}
where $\rho_c$ is the critical density and $h$ is the normalized Hubble parameter. DM represents the 26.6\% energy density of the universe. Despite this precise measurement, the fundamental nature of DM remains one of the most important unsolved mysteries in modern cosmology. 
\subsection{Dark matter as a particle}
One of the leading paradigms to explain the nature of DM, is that it consists of a new particle that couples very weakly with Standard Model (SM) particles. Assuming DM to be a new kind of particle, it could have been generated thermally in equilibrium alongside the baryonic matter from annihilation processes within the primordial plasma. In the simplest scenario, a DM particle was initially in thermal equilibrium and decoupled as the Universe expanded and cooled, a process known as freeze-out. In this case, the present-day abundance depends primarily on the particle’s annihilation cross section \cite{CerdenoAstro2025}. Remarkably, from particle physics and cosmological calculations, the required value for the cross section for a particle with mass in the GeV–TeV range to match the relic abundance gives \(\langle \sigma v \rangle \approx 3 \times 10^{-26}~\mathrm{cm^3\,s^{-1}}\). This value represents, in fact, interactions of the order of the weak force \(\langle \sigma v \rangle \sim G_F^2\, m_{\mathrm{WIMP}}^2\) where the relic abundance naturally matches the observed DM density. This coincidence, often refereed as the \emph{WIMP Miracle}, motivated decades of experimental searches for Weakly Interacting Massive Particles (WIMPs), including direct detection, indirect detection, and collider experiments.

\subsubsection{Sub-GeV dark matter}
Although (WIMPs) have long been considered the leading candidates in particle DM scenarios, the lack of evidence from experimental searches, together with new astrophysical observations, has motivated the exploration of alternative possibilities. Indirect searches with instruments such as Fermi-LAT \cite{Atwood_2009}, H.E.S.S \cite{hinton2004status}. or AMS-02 \cite{aguilar2013first}  have not yet given definitely signals. Although WIMPs could annihilate into photons, giving $\gamma$-rays signatures, the apparent Galactic Center $\gamma$-ray excess is now more likely attributed to unresolved astrophysical sources like mili-second pulsars \cite{bartels2016strong}. Also, stringent bounds from antiprotons and dwarf spheroidal galaxies exclude large portions of the classic WIMP parameter space \cite{gaskins2016review, luque2024antiprotonboundsdarkmatter}. On the other hand, direct detection experiments have reached impressive sensitivities, probing interaction cross sections close to the so-called neutrino floor, which represents an irreducible background from solar and atmospheric neutrinos were it would be hard to disentangle between DM scattering signals and those of the neutrinos \cite{gaskins2016review}. Out of the WIMP scenario, one particularly well motivated and largely unexplored mass range lies well below the GeV scale: the so called sub-GeV DM. This region is highly motivated by several astrophysical anomalies, such as the 511 keV excess observed in the Galactic Center \cite{kierans2020detection}, the diffuse excess of gamma rays in the MeV band reported by instruments like INTEGRAL/SPI \cite{STRONG_2011, orlando2018imprints}, and the apparent over-ionization of molecular gas in the Central Molecular Zone (CMZ), which cannot be easily explained by standard CR populations \cite{PhysRevLett.134.101001}. This regime exhibits a rich phenomenology, opening up a variety of potential production mechanisms, interactions, and detection channels. Unlike the traditional WIMP paradigm, it no longer relies on weak interactions but instead points toward a distinct dark or hidden sector beyond the SM. In the MeV mass regime, several hidden sector scenarios remain viable, where the DM particle interacts with the SM through specific portals. This is because they evade the stringent limits placed on WIMPs coming from indirect detection on astrophysical observables and direct detection experiments, which have energy thresholds too high to detect them. Well motivated models, among many more, could be:

\begin{itemize}
    \item \emph{Vector portals}, in which a light dark photon mixes kinetically with the SM hypercharge, which allows thermal freeze out (produced at equilibrium and decoupling later from the plasma) or freeze in (produced already out of the equilibrium) \cite{galison1984two, holdom1986two}. It presents small couplings that arise from loop effects.
    \item \emph{Scalar portals}, where a light scalar mixes with the Higgs boson. It can naturally accommodate particles in the MeV mass range \cite{patt2006higgs}, although some models face strong meson decay constraints \cite{Krnjaic_2016}. 
    \item \emph{Pseudoscalar portals}, where DM couples to axion-like or pseudoscalar mediators. Such models naturally lead to velocity-suppressed annihilation channels (often p-wave dominated, which means the process is mainly governed by the partial wave with orbital angular momentum l=1, leading to a cross section proportional to $v^2$). This helps to evade CMB and Big Bang nucleosynthesis constraints, while still permitting a viable thermal history. In addition, pseudoscalar interactions show a rich phenomenology at the MeV–GeV scale, including possible connections to the 511 keV excess in the Galactic Center and distinctive indirect detection signatures \cite{aghaie2025halping511kevline}.
\end{itemize}
Even \emph{Strongly Interacting Massive particles (SIMP)} \cite{battaglieri2017us} are allowed, at which the dark sector is confined and the masses of the DM particles are within the MeV regime $5\leq m_\chi \leq 200$ MeV. As it can be seen, there is a lot of permitted models that represent a compelling DM candidate, even though they need to invoke a new dark sector to keep matching the relic abundance.\\

If it is wanted to consider a thermal production, just as in the WIMP scenario, that retrieves the correct relic abundance, it can be differentiated two situations \cite{battaglieri2017us}: DM can annihilate either into lighter mediators \textbf{(secluded regime)} or \textbf{directly} into SM fermions via an s-channel portal, where the mediator is produced in the intermediate state. The latter case offers a predictive target since thermal history imposes a minimum coupling between the mediator and the SM. However, strong constraints from CMB measurements require that late time annihilation rates be highly suppressed \cite{finkbeiner2012searching}, favoring models with p-wave processes, inelastic channels, or particle–antiparticle asymmetries. In the asymmetric scenario, the DM relic density is set not only by annihilation but also by an initial particle–antiparticle imbalance. As a consequence, today’s annihilation rate can be drastically reduced, since only the small symmetric component survives, while still achieving the correct abundance \cite{Lin_2012}. This feature of asymmetric DM is particularly interesting, as it reproduces the correct relic abundance and avoids the stringent limits placed from indirect detection on symmetric scenarios \cite{Balan_2025}. DM could also decay into SM particles, although the DM must be sufficiently long lived to keep its cosmic density almost constant until nowadays \cite{CerdenoAstro2025}. These conditions leave a good and a viable, testable parameter space for future direct detection, indirect detection and accelerator searches. \\

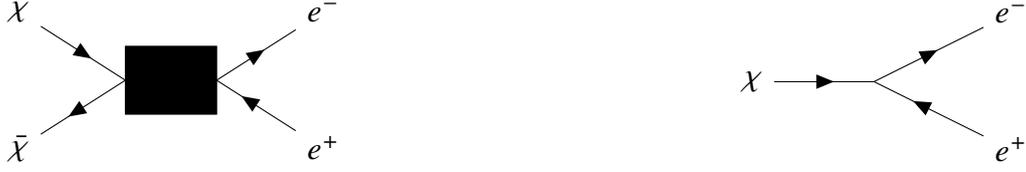
\begin{figure}[h]
\centering

\begin{subfigure}[t]{0.45\textwidth}
\centering
\begin{tikzpicture}
\begin{feynman}
  \vertex (i1) at (-2, 0.9) {\(\chi\)};
  \vertex (i2) at (-2,-0.9) {\(\bar{\chi}\)};
  \vertex (f1) at ( 2, 0.9) {\(e^{-}\)};
  \vertex (f2) at ( 2,-0.9) {\(e^{+}\)};
  \vertex (L) at (-0.6, 0);
  \vertex (R) at ( 0.6, 0);

  \diagram*{
    (i1) -- [fermion]      (L),
    (i2) -- [anti fermion] (L),
    (R)  -- [fermion]      (f1),
    (R)  -- [anti fermion] (f2),
  };
\end{feynman}
\node[rectangle, fill=black, draw, minimum width=12mm, minimum height=9mm] at (0,0) {};
\end{tikzpicture}
\caption{DM annihilation into an electron--positron pair, represented as an effective interaction (black box).}
\end{subfigure}
\hfill
\begin{subfigure}[t]{0.45\textwidth}
\centering
\begin{tikzpicture}
\begin{feynman}
  \vertex (i)  at (-1.6,0) {\(\chi\)};
  \vertex (v)  at (0,0);
  \vertex (o1) at (1.8, 0.9) {\(e^{-}\)};
  \vertex (o2) at (1.8,-0.9) {\(e^{+}\)};

  \diagram*{
    (i) -- [fermion] (v),
    (v) -- [fermion] (o1),
    (v) -- [anti fermion] (o2),
  };
\end{feynman}
\end{tikzpicture}
\caption{DM decay into an electron--positron pair through a single vertex.}
\end{subfigure}

\caption{Representative Feynman diagrams for DM interactions producing electrons and positrons. (a) Effective annihilation channel \(\chi\bar{\chi}\to e^- e^+\). (b) Direct decay channel \(\chi \to e^- e^+\).}
\label{feynmann}
\end{figure}

In this work, the focus will not be on a specific model, but rather on their phenomenological consequences. The approach is therefore agnostic regarding the underlying processes and instead aims to study the observable implications of light DM. By concentrating on the phenomenology of annihilation and decay channels, and the resulting astrophysical signatures that could naturally explain the anomalies. This analysis remains broadly applicable across a wide class of models, without committing to a particular theoretical framework. More precisely, it is considered the effective interactions shown in Figure (\ref{feynmann}) with DM particles annihilating and decaying into electrons and positrons. 

These low-energy particles are expected to strongly ionize $H_2$. This is why it may be interesting to search for their footprint in certain astrophysical objects, such as molecular clouds, as will be shown below.
\subsection{Primordial black holes as a candidate}
An alternative to the particle DM paradigm is the hypothesis that DM could be composed of macroscopic objects. Most of these possibilities have already been ruled out by observations, leaving one: PBHs, as one of the most compelling remaining candidate. PBHs are black holes formed in the earliest stages of the Universe without the requirement of the collapse of a star. They could still play a relevant role even if they do not account for the entire DM density, within a multicomponent scenario. Indeed, ordinary astrophysical black holes already contribute to a small fraction of the measured DM. The pioneering works by Hawking and Carr \cite{hawking1971gravitationally, carr1974black} first pointed out that such objects could form from density fluctuations in the primordial Universe. More recently, PBHs have been shown to naturally arise in many inflationary models (see, e.g. \cite{garcia1996density, carr1993primordial}). Following these discoveries, PBHs began to be considered as viable DM candidates. Recalling the conditions said at the beginning of this section (non baryonic, no interaction with light, etc.) PBHs appear as a possible explanation for DM, without invoking the existence of new fundamental particles.\\ 

\begin{figure}[H]
    \centering
    \includegraphics[width=0.6\linewidth]{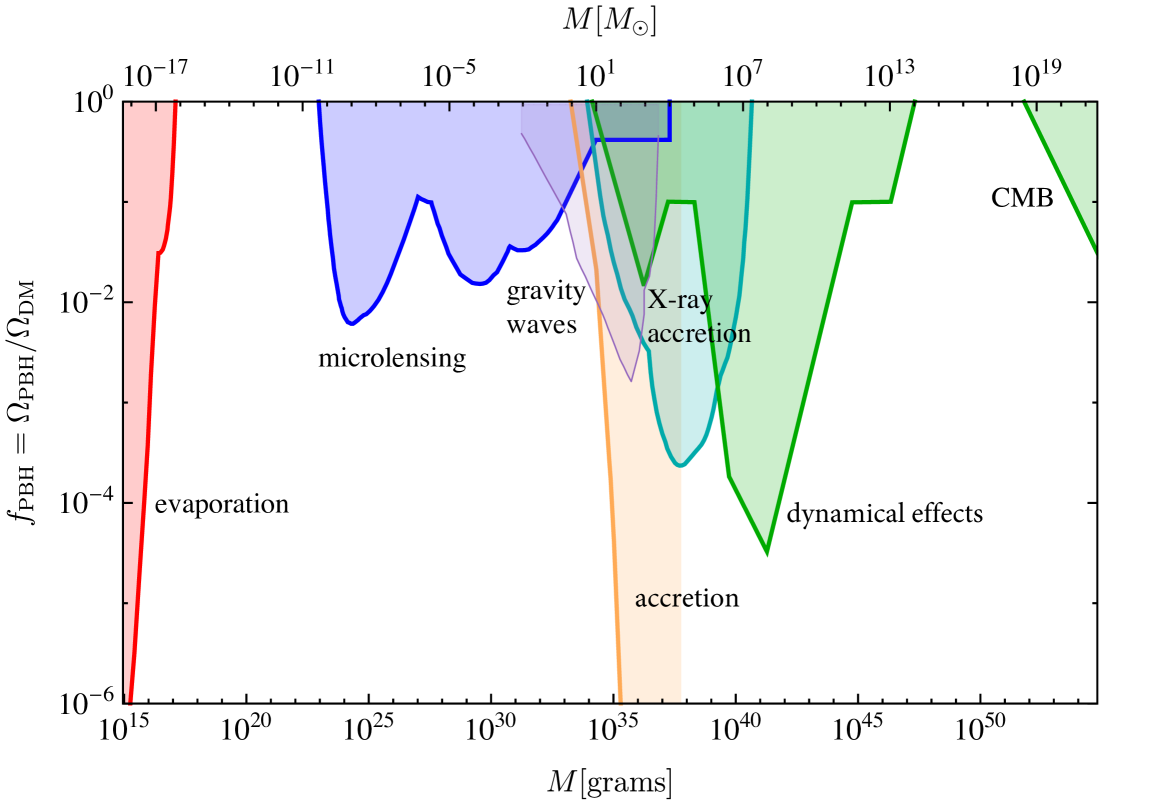}
    \caption{Main constrains on the fraction of DM $f_{PBH}$ as PBHs as a function of its mass \cite{carr2025primordialblackholes} }
    \label{pbhconstrainsfigure}
\end{figure}

For many years, the scientific community considered black holes to be eternal. However, earlier studies suggested that PBHs could have very small masses \cite{zel1966hypothesis, hawking1971gravitationally}. This insight motivated Stephen Hawking to analyze the behaviour of quantum fields in curved spacetime, and ultimately led him to the incredible discovery that black holes could emit thermal radiation, thereby losing mass and energy over time \cite{hawking1974black, hawking1975particle}. This process, now known as Hawking radiation, represents the power (energy per unit time) emitted by a black hole as a function of their temperature $P \propto T^2$. The associated particle flux also depends on the temperature, following approximately $\dot{N} \propto T$ \cite{hawking1975particle}. The exact temperature depends on the type of the BH. For instance, for Schwarzschild black holes (non rotating, non charged) the temperature associated scales with mass as:
\begin{equation}
T_{\rm BH} \;\approx\; \frac{\hbar\,c^{3}}{8\pi\,G\,M\,k_{B}}
\;\simeq\;
10\ \mathrm{MeV}\,\times\!\biggl(\frac{10^{20}\,\mathrm{g}}{M}\biggr)
\end{equation}
It follows that lower-mass PBHs are hotter and radiate more intensely, which imposes a lower limit on the mass of any PBH that could still exist today. PBHs formed in the early Universe must have had an initial mass greater than $M_{min} = 7.5\cdot10^{14}$g \cite{luque2024refininggalacticprimordialblack} in order to survive without being fully evaporated at the present time. Evaporating PBHs are of particular interest because their particle emission can be used to place constraints on their present-day abundance. The method followed in this work (see Section 4) is analogous to how decays or annihilations of particle DM are used to constrain its decay or annihilation rates, although in the PBH case, the emission rate is determined by the mass of the black hole rather than by an interaction cross section or a decay lifetime. This work is restricted to the conservative case of monochromatic and spinless PBHs, i.e., it is assumed a single mass instead of a mass distribution and angular momentum is neglected. Figure (\ref{pbhconstrainsfigure}) shows the current bounds on PBHs of different masses. In particular, PBHs with asteroid-like masses $10^{14}-10^{23}$g are very interesting because there are no constrains below $f_{PBH} = 1$ (where $f_{PBH}$ represents the ratio of DM as the form of PBHs) on most of this range. It could therefore constitute a great percentage of DM, if not all of it. In this case of asteroid-mass PBHs, a similar emission of comparable energies is also expected as in the particle scenario, which could likewise leave an imprint in molecular clouds (Section 3).

\newpage
\section{Expected emissions from dark matter}
The DM candidates mentioned above present an interesting phenomenology and are expected to generate emissions with observable consequences in the interstellar medium that can be measured, either DM is a novel particle or macroscopic objects such as PBHs. Such emissions can be modeled through source functions usually denoted by $Q$, which represent the rate of injected particles (per unit volume, time, and energy) and whose functional form depends on the underlying nature of DM.

\subsection{Dark Matter particle emissions}
For annihilating and decaying DM particles into electrons and positrons:

\begin{equation}
Q_{e}\bigl(\vec{x},E_{i}\bigr)
=
\begin{cases}
\dfrac{\langle\sigma v\rangle}{2}\,\biggl(\dfrac{\rho_{\chi}(\vec{x})}{m_{\chi}}\biggr)^{2}\,\dfrac{\mathrm{d}N^{\rm ann}_{e}}{\mathrm{d}E_{e}}
& (\text{annihilation})\,,\\[1em]
\Gamma\,\biggl(\dfrac{\rho_{\chi}(\vec{x})}{m_{\chi}}\biggr)\,\dfrac{\mathrm{d}N^{\rm dec}_{e}}{\mathrm{d}E_{e}}
& (\text{decay})\,.
\end{cases}
\end{equation}
where $\langle\sigma v\rangle$ is the thermally averaged cross section and $\Gamma = \frac{1}{\tau}$ is the decay rate defined as the inverse of the decay lifetime $\tau$, $\rho_\chi$ is the DM density of the region, $m_\chi$ the mass of the DM particle and $\tfrac{dN}{dE}$ is the injection spectrum, which gives the number of electrons and positrons produced per DM process. This quantity defines the source term, but for the purposes of the simulations the relevant observable will be the differential flux, $\tfrac{d\phi_e}{dE}$, introduced below. The flux is directly related to the source function through Eq. (\ref{difflux}), and its explicit form will be discussed together with the spectra shown further down (Section 4).

\subsection{Asteroid-Mass Primordial Black Holes emissions}
The source function of PBH can be modeled similarly to Eq (3) of \cite{luque2024refininggalacticprimordialblack}. In this work, however, it is expressed as:

\begin{equation}
Q_{\rm PBH}(E)
= n_{\rm PBH}\,\cdot\,\frac{dN}{dtdE}\bigl(m_{\rm PBH},E\bigr)\,
\label{sourcePBH}
\end{equation} where the PBH number density is
\begin{equation}
n_{\rm PBH}
= f\,\frac{\rho_{\rm DM}}{m_{\rm PBH}}\,
\end{equation}
with f denoting the fraction of DM in the form of PBHs and $\rho_{DM}$ the DM density distribution adopted also in the particle DM scenario in order to maintain consistency.\\

\begin{figure}[H]
\begin{subfigure}[b]{0.5\textwidth}
    \centering
    \includegraphics[width=\linewidth]{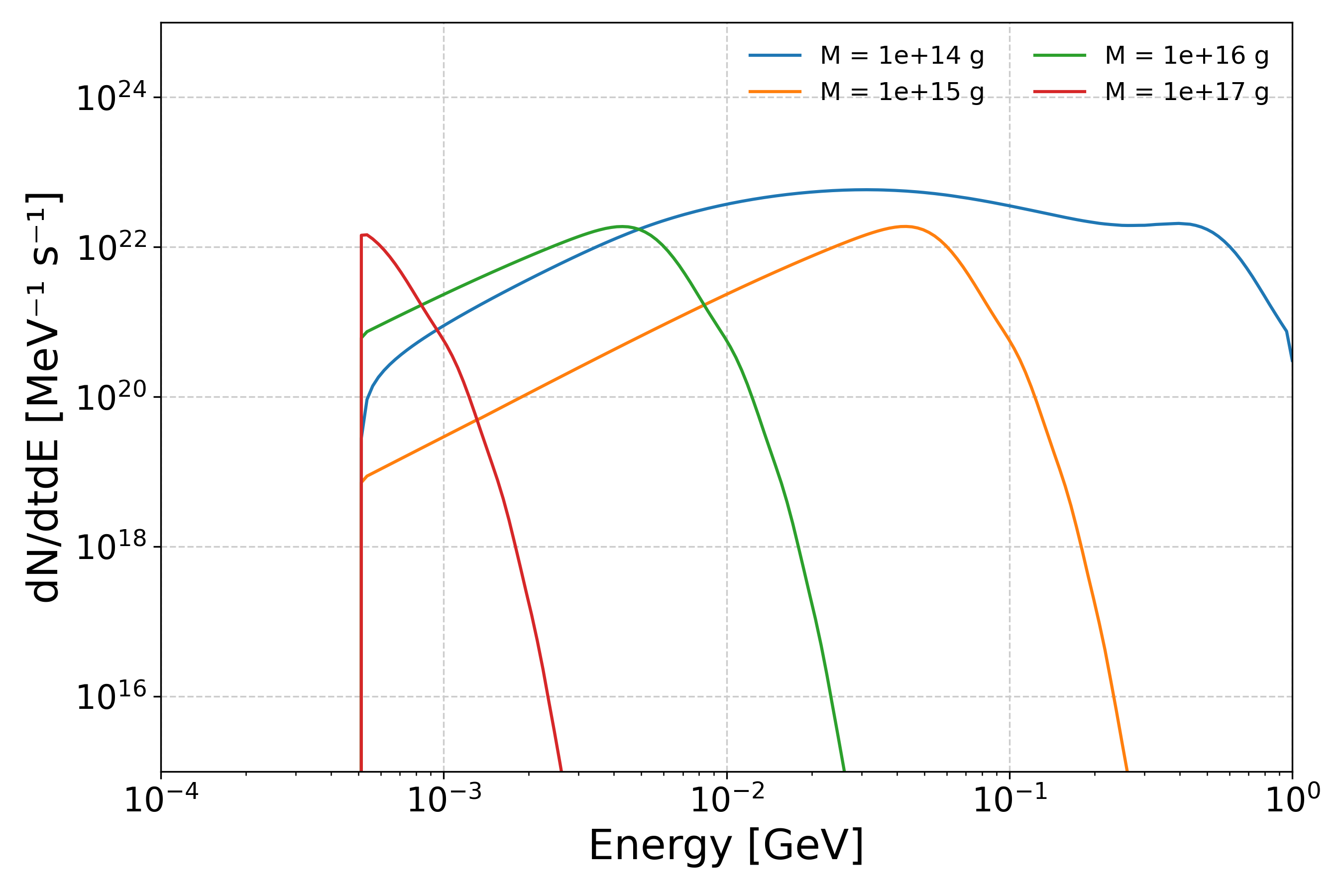}
    \caption{Electron's injection rate}
    \label{inyrateelect}
\end{subfigure}
\hfill
\begin{subfigure}[b]{0.5\textwidth}
    \centering
    \includegraphics[width=\linewidth]{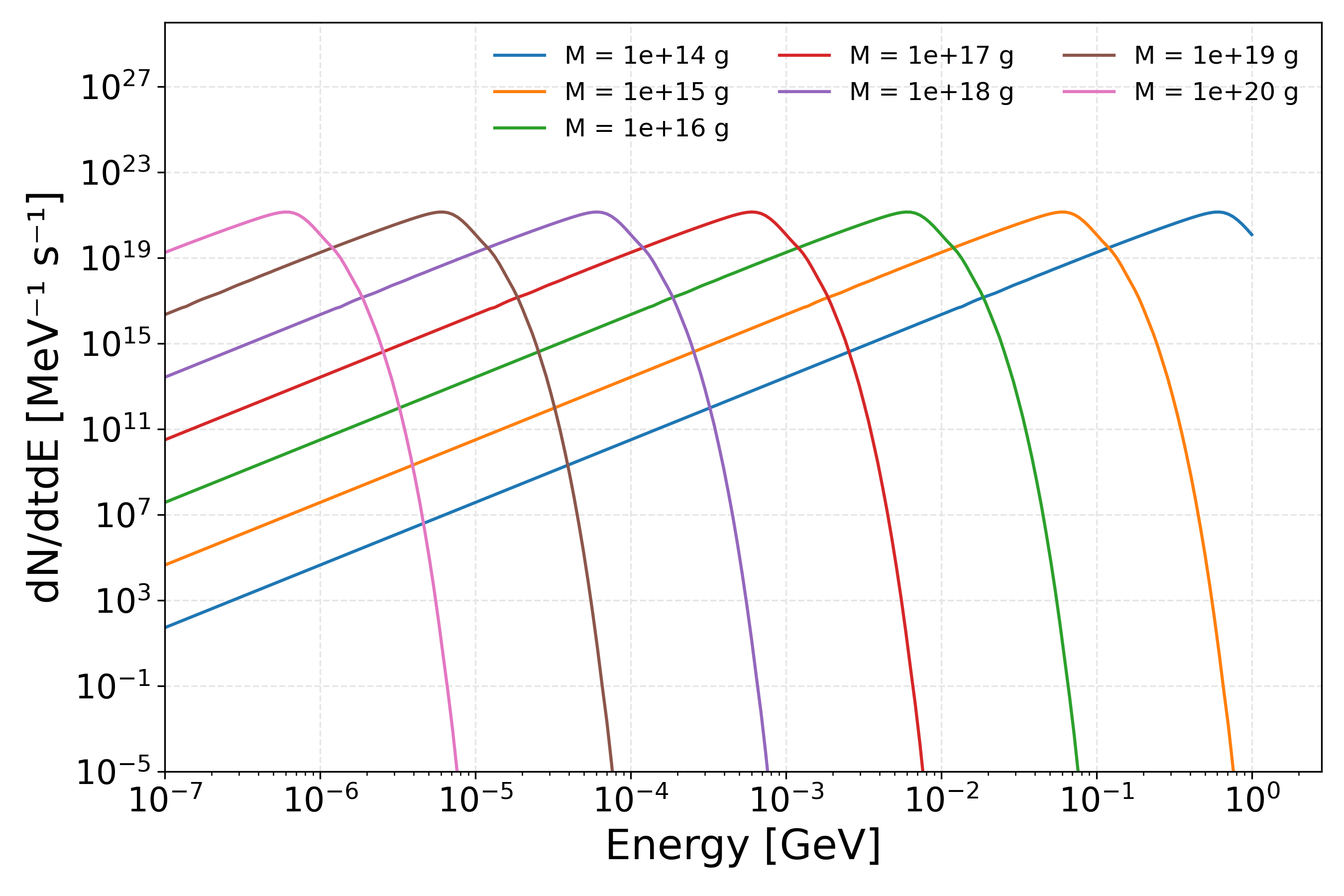}
    \caption{Photon's injection rate}
    \label{inyratepho}
\end{subfigure}
\caption{Injection rate of electrons and photons coming from Hawking's radiation in Schwarzschild asteroid-mass PBHs as a function of its energy for various masses}
\label{inyrateeandpho}
\end{figure}

The emission spectrum $\frac{dN}{dtdE}$ is obtained from a code named \emph{BlackHawk} \cite{Arbey_2019}. BlackHawk is a public C code (Linux-based) designed to compute the Hawking radiation spectra of arbitrary distributions of Schwarzschild and Kerr black holes. It solves the spin and energy dependent transmission probabilities (“greybody factors”) for each particle species. Then, it integrates the resulting emission spectra over time giving hadronized particles outputs. The code also allows for the inclusion of particles beyond the Standard Model, as well as the treatment of extended PBH populations. In the context of this work, only the instantaneous spectra of primary and secondary emissions for a single Schwarzschild black hole is required, corresponding to the most conservative scenario. That way, the  interpolated spectrum from BlackHawk is used as input in the equation (\ref{sourcePBH}) including only photon and $e^+e^-$ channels. For asteroid mass PBHs, BlackHawk yielded the spectra results shown in Figure (\ref{inyrateeandpho}).

There are no emissions of electrons and positrons above $\sim10^{17}$g for Schwarzschild black holes. This was the expected result, since a black hole beyond that mass has a temperature $T < 100$ keV, well below the energy of an electron at rest $\sim 0.511$ MeV. In the photon case, there will always be emission, but the spectrum peaks at lower energies as the black hole mass increases. Consequently, the most efficient sources of ionizing photons are the lighter black holes, whose higher Hawking temperatures shift the peak to the MeV–GeV range. However, for this range of energies, the photon contribution is insignificant and therefore neglected \cite{Luque_2025}.
\newpage

\section{Ionization rates of Molecular Clouds as the observable}
Molecular clouds (MCs) are among the densest, coldest, and most chemically complex objects in the interstellar medium (ISM). They are the primary sites of star formation, as the gravitational collapse of their dense cores can lead to the birth of new stars. Their composition is dominated by molecular hydrogen, with significant amounts of helium and trace elements such as carbon, oxygen, and nitrogen. The prevalence of molecules in these environments is favored by their relatively low temperatures ($T \sim 10$K), which suppress thermal dissociation rates. Nevertheless, molecular formation requires an external source of energy, such as Cosmic Rays (CRs) or ultraviolet (UV) photons, to drive endothermic reactions or desorption from dust grains. MCs are also magnetized, with field strengths typically ranging from a few $\mu$G to several mG. The degree of magnetization depends on the local ionization fraction, although its modeling is not yet well understood. In many cases, this value is higher than what would be expected from purely thermal ionization at such low temperatures. In addition, turbulence inside the clouds, observed through the broadening of spectral lines, acts as an additional source of support, partly counteracting gravitational collapse \cite{gabici2022low}.\\

A convenient way to classify MCs is through their hydrogen 
column density, $N_{\mathrm{H}_2}$. Diffuse MCs, with 
$N_{\mathrm{H}_2} \sim 10^{20} - 10^{22}\,\mathrm{cm}^{-2}$, are relatively 
transparent to UV radiation. In these environments, hydrogen is partly 
molecular and carbon is mainly in the form of C$^+$, leading to a typical 
ionization fraction of $x_e \sim 10^{-4}$. By contrast, in dense/ dark MCs with $N_{\mathrm{H}_2} \gtrsim 10^{22}\,\mathrm{cm}^{-2}$, 
UV photons are strongly attenuated and absorbed in the outer layers. 
Hydrogen is almost entirely molecular (H$_2$), carbon is locked in CO, 
and the ionization fraction drops to $x_e \sim 10^{-7}$. In this regime, 
low-energy CRs ($E < 1\,\mathrm{GeV}$) are expected to dominate 
the ionization processes within the cloud \cite{gabici2022low}.\\

The ionization rate ($\zeta$) in MCs can be measured indirectly 
through the abundance of the molecular ion H$_3^+$. The classical reaction 
chain starts with the ionization of H$_2$:

\begin{align}
\mathrm{H}_2 + \mathrm{CR} &\longrightarrow \mathrm{H}_2^+ + e^-,\label{reaction1} \\[6pt] 
\mathrm{H}_2^+ + \mathrm{H}_2 &\longrightarrow \mathrm{H}_3^+ + \mathrm{H}
\label{reaction2}
\end{align}
where CR stands for electrons, protons, etc. However, in our modeling the particles coming form the DM sources would be entirely electrons, positrons and photons in the PBH case. The ionization rate can then be obtained by the following relation:
 \begin{equation}
     \zeta_{H_2}\,n(\mathrm{H}_2) = k_e\,n(\mathrm{H}_3^+)\,n_e
 \end{equation}
 where $k_e$ is the H$_3^+$ electron recombination rate. By measuring the column density of H$_3^+$ via its infrared absorption lines, it can inferred $\zeta_{H_2}$ accurately. These measurements are commonly complemented by other tracers including the millimeter‐wave ions HCO$^+$, DCO$^+$ (sensitive to $\zeta$ and temperature in “hot cores”), and hydroxyl (OH) or HD in more diffuse clouds. It is important to search for several tracers since, for example, in really dense clouds, H$_3^+$  is mostly destroyed
 through the proton hop reaction \cite{gabici2022low}
 \begin{equation}
     \mathrm{H}_3^+ + \mathrm{CO} \longrightarrow \mathrm{HCO}^+ + \mathrm{H}_2
 \end{equation}

\begin{figure}[H]
    \centering
    \includegraphics[width=0.5\linewidth]{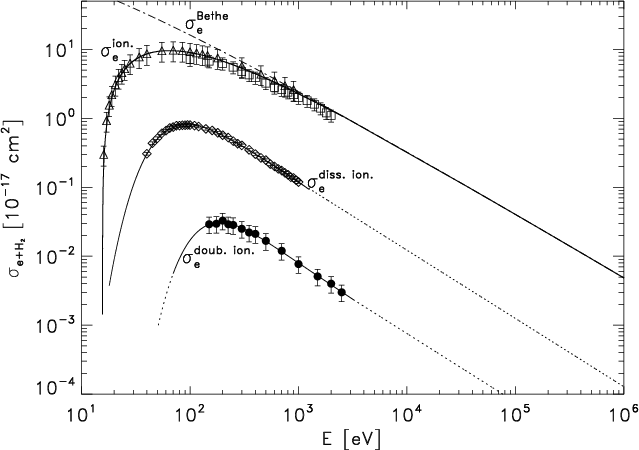}
    \caption{Electron impact cross sections for H$_2$—single ionization ($\sigma_e^{\rm ion}$), dissociative ionization ($\sigma_e^{\rm diss\,ion}$), and double ionization ($\sigma_e^{\rm doub\,ion}$)—as a function of its energy. They are plotted as polynomial fits to the solid portions of the curves in the table 2 of \cite{padovani2009cosmic}. The dot–dashed line shows the Bethe single‐ionization cross section scaled by a factor of two. Experimental data symbols (triangles, squares, diamonds, filled circles) are those compiled and discussed in that review.}
    \label{crosssectionCRMC}
\end{figure}

Ionization rates measured in both diffuse and dense MCs are unexpectedly high, often exceeding by one to two orders of magnitude the values predicted from the local CR spectrum, suggesting the presence of an additional low-energy ionizing component \cite{padovani2009cosmic, Ravikularaman_2025}. Yet the origin of these enhanced rates remains unclear, the interpretation is challenged by significant uncertainties arising both from the modeling of CR transport at sub-GeV energies and from the observational diagnostics themselves. In this context, a population of MeV-scale DM particles, either annihilating or decaying into 
electrons and positrons within the cloud, or alternatively primordial black holes evaporating via Hawking radiation, could provide an additional ionization component and thereby offer a natural explanation for the observed excess.  The resulting $e^\pm$ with energies in the keV–MeV range have cooling lengths comparable to cloud sizes and efficiently ionize H$_2$ via the classical reaction chain of equations (\ref{reaction1})-(\ref{reaction2}).  As a result, the total ionization rate
\[
\zeta_{H_2}^{\rm (tot)} \;=\; \zeta_{H_2}^{\rm (CR)} \;+\; \zeta_{H_2}^{\rm (DM\;\!e^\pm)}
\]
can reach the observed values without extra effects.  Moreover, since the DM induced $e^\pm$ injection is insensitive to cloud column density above $N_{H_2}\sim10^{22}\rm\,cm^{-2}$, this mechanism also helps explain why dark clouds exhibit similarly enhanced ionization rates despite their strong UV shielding. \ref{crosssectionCRMC}

In fact, as suggested by this figure, the efficiency of interaction between $\text{e}^+\text{e}^-$ and $H_2$ reaches its peak at really low energies $\sim 100 \text{eV}$. For MeV DM particles decaying and annihilating into electrons and positrons and electron emissions from asteroid-mass PBHs, it can be assumed that all of their energy is deposited within the MC. It is necessary to highlight that these DM processes are not only produced outside of the MC, but within it. In fact, the relevant values of the ionization caused by the DM are those that are produced inside the cloud, since they avoid external energy losses and let each particle traverse the full dense path, maximizing primary ionization. This processes will also form secondary electron cascades amplifying the effect, yielding a uniformly high ionization rate compared to externally CRs.

\newpage
\section{Method and theoretical framework}

The transport equation of CRs inside a MC can be expressed as \cite{Shi_2024}:
\begin{equation}
    \frac{\partial N}{\partial t}
- \nabla\!\cdot\!\bigl(D\,\nabla N\bigr)
- \frac{\partial}{\partial p}\!\bigl(b\,N\bigr)
= Q
\label{TransEq}
\end{equation}
Here, the quantity
\[
N(\mathbf{r},p,t) \;=\; 4\pi\,p^{2}\,f(\mathbf{r},p,t)
\]
represents the cosmic‐ray number density per unit momentum, with \(f(\mathbf{r},p,t)\) denoting the isotropic phase‐space distribution function.  The coefficient 
\(\,D(\mathbf{r},p)\,\) is the spatial diffusion constant (assumed isotropic), while 
\[
b(\mathbf{r},p)\;=\; -\frac{\mathrm{d}p}{\mathrm{d}t}
\]
gives the rate at which particles lose momentum.  Finally, \(Q(\mathbf{r},p,t)\) specifies the source term for particle injection. Because the gas density inside MCs varies from place to place, both the diffusion coefficient \(D\) and the momentum‐loss rate \(b\) acquire an explicit dependence on position \(\mathbf{r}\). Equation (\ref{TransEq}) has no analytical solution, except in the case of a uniform gas distribution, which is the assumption adopted in this work for simplicity..\\

The method implemented here is based on evaluating the ionization rate caused by DM emission in selected MCs and comparing it with the measured values, in order to determine the DM annihilation cross section $\langle\sigma v\rangle$, the decay lifetime $\tau$, or the fraction of DM in the form of PBHs, $f_{\text{PBH}}$, that are compatible with the observations. In order to set a conservative approach, it is assumed that the entire observed ionization rate in the MCs originates from DM-induced ionizations. For that task, the ionization rate has to be computed for the expected emissions of DM, which can be done with the following expression:

\begin{equation}
\zeta \;=\; 2 \cdot 4\pi 
\int_{E_{\min}}^{E_{\max}}
J(E, x)\,\sigma(E)\,\bigl(1 + \theta_{e}(E)\bigr)\,\mathrm{d}E
\end{equation}
where the factor 2 accounts for the two particles produced in the particle scenario (not necessary for the PBH case),  $E_{max}$ is the maximum kinetic energy the electrons and positrons can carry and $E_{min}$ the minimum energy required in order to produce an ionization on hydrogen (13.6eV). $\sigma(E)$ is the cross section between $e^+e^-$ and $H_2$ molecules calculated as in Eq.(7) of \cite{padovani2009cosmic}, $\theta_e$ denotes the average number of secondary ionizations produced by each primary ionization event (Eq. (2.23) of \cite{krause2015crimecosmicray}) and $J(E,x)$ is the J function which represents the electron current density function and has the form:

\begin{equation}
J(E,\mathbf{x}) \;=\; \frac{\mathrm{d}\phi_e}{\mathrm{d}E}(E,\mathbf{x}) \;\frac{\beta_e(E)\,c}{4\pi}
\end{equation}
Here, $\beta_e$ stands for the speed of the electron or positron and $\frac{\mathrm{d}\phi_e}{\mathrm{d}E}$ is their differential flux inside the clouds. It can be expected that this injection of $e^+e^-$ would produce ionizating photons. However, as mentioned in the PBH emission section, those are neglected, since at this range of energies their cross-section is about $10^{-6}$ times lower that the electrons and positrons. Hence, no relevant ionization will be expected \cite{Luque_2025}. The analytical approximation made for the code consists on neglecting the diffusion coefficient within the cloud and the differential flux at steady-state is computed as:\\

\begin{equation}
\frac{\mathrm{d}\phi_e}{\mathrm{d}E}(E,\mathbf{x})
=
\frac{1}{b_e\bigl(E,n(\mathbf{x})\bigr)}
\int_E^{\infty} Q_e\bigl(E',\mathbf{x}\bigr)\,\mathrm{d}E'
\label{difflux}
\end{equation}

This approximation assumes that transport inside the MC is dominated by continuous energy losses, while spatial diffusion plays a subdominant role. Neglecting diffusion is therefore justified for compact, dense clouds and/or for low-energy CRs, where the energy loss timescale of sub-GeV electrons is much shorter than their diffusive escape time. However, in more extended environments and with high energetic particles, at which more efficient propagation is allowed, diffusion becomes larger and cannot be neglected. The approximation behaves well up to a few tens of MeV and helps simplifying the calculations. Now, the loss term is defined as $b_e = \frac{dE^{ion}}{dt}$ calculated as shown in Eq. (C.39) of \cite{Evoli_2017} and $Q_e$ will vary depending on the different scenarios described on section 2.\\

\begin{figure}[h!]
     \begin{subfigure}[b]{0.5\textwidth}
         \centering
         \includegraphics[width=\linewidth]{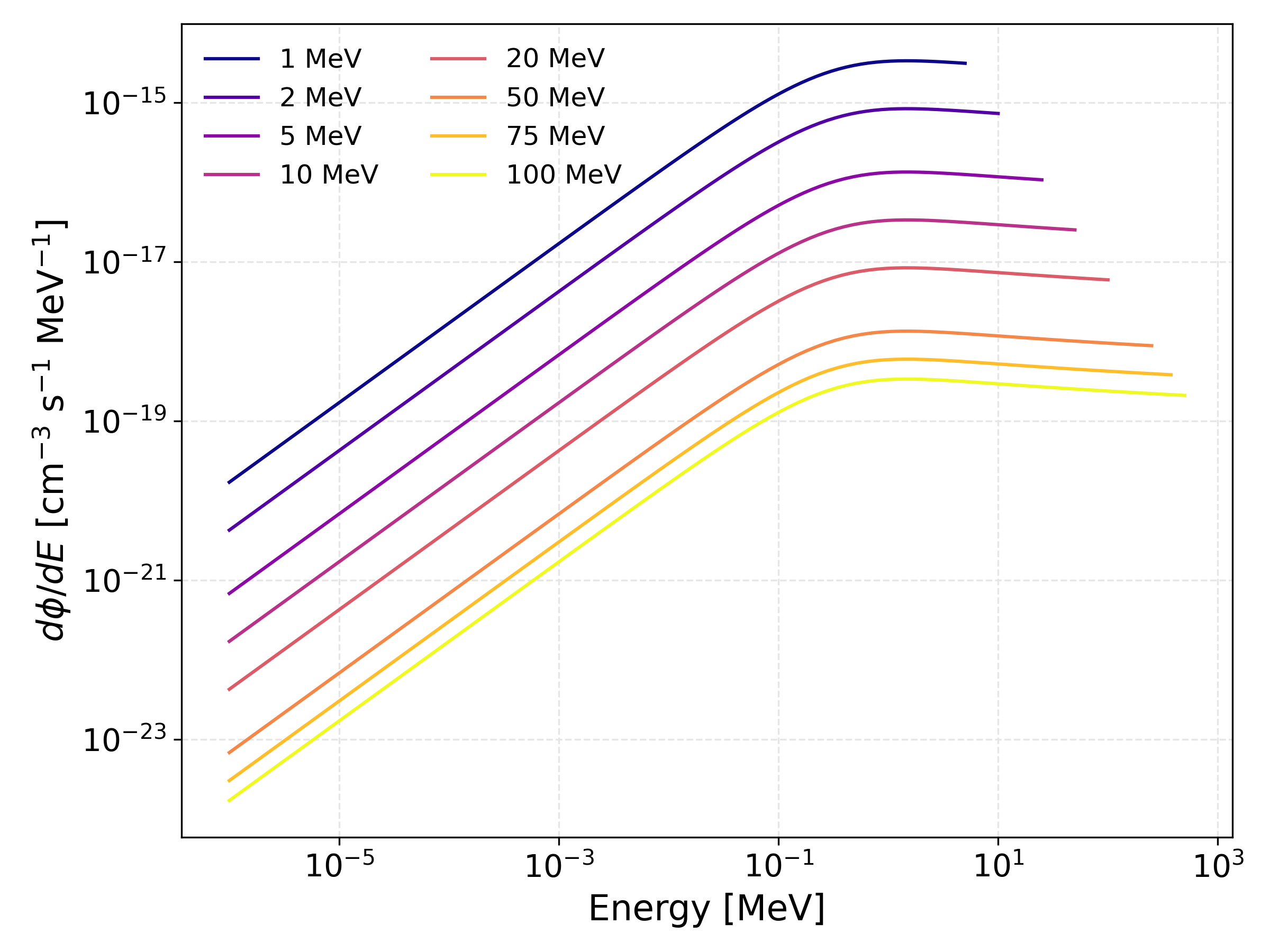}
         \caption{Annihilation spectra}
         \label{annihspec}
     \end{subfigure}
     \hfill
     \begin{subfigure}[b]{0.5\textwidth}
         \centering
         \includegraphics[width=\linewidth]{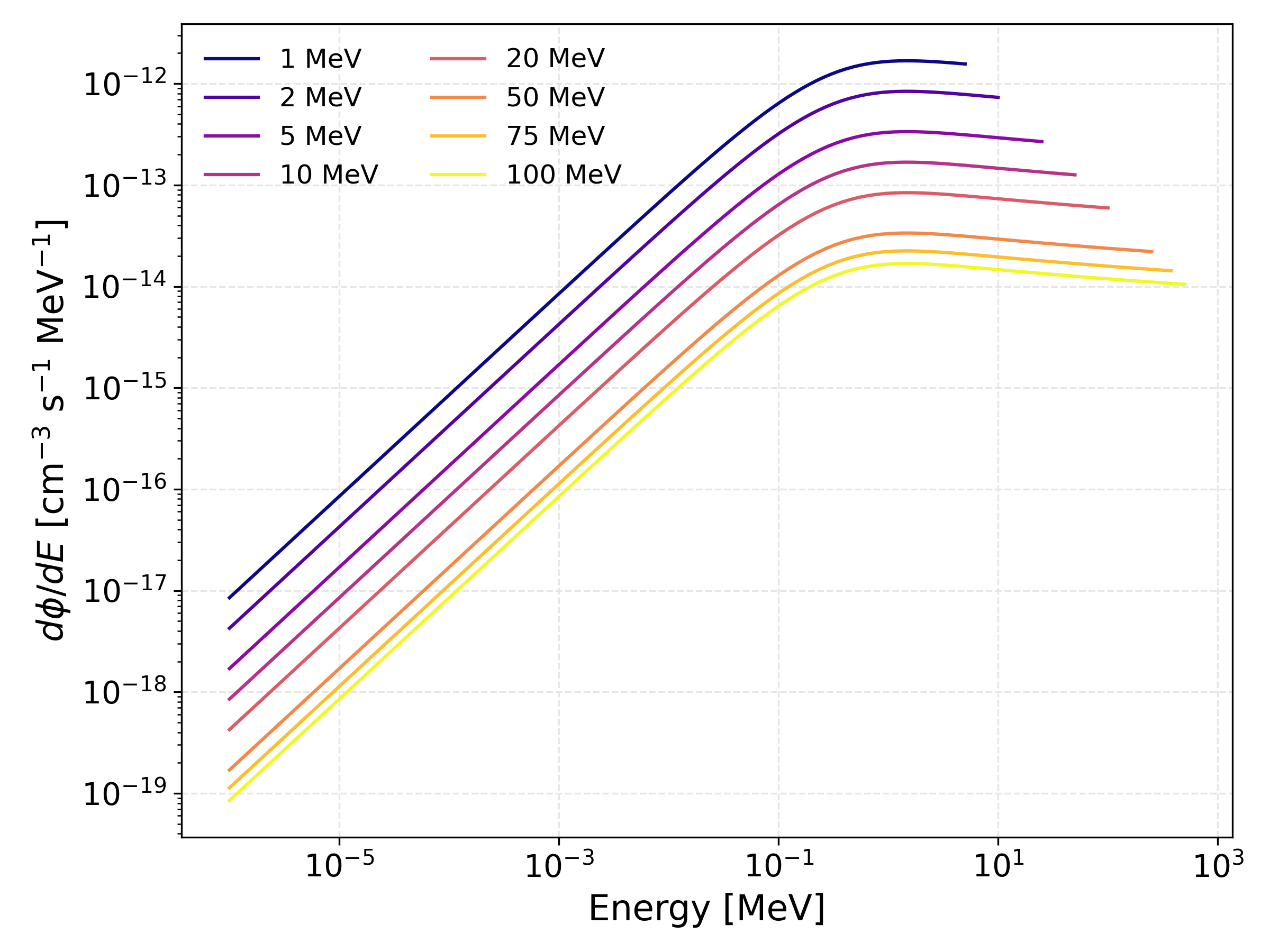}
         \caption{Decay spectra}
         \label{decspec}
     \end{subfigure}
    
        \caption{Differential flux of electrons and positrons of annihilating and decaying DM as a function of its energy for various masses}
        \label{MCIR}
\end{figure}

As can be seen, the differential flux is directly proportional to the source function. Now that it is defined, recalling the DM particle scenario, the differential flux computed can be seen in Figure (\ref{MCIR}). In the numerical implementation developed for this work, 
it is not necessary to evaluate the full integral in Eq. (\ref{difflux}). Instead, the source term 
$Q_e$ is treated as a set of discretized Dirac deltas for each DM mass under consideration, 
$m_\chi = 1, 2, 5, 10, 20, 50, 75, 100~\text{MeV}$. However, the resulted flux follows a continuum spectrum for each particular mass. This is related to the energy loses term $b_e(E, n(x))$ which redistributes the particles over different energies. Otherwise, the flux would reduce to a Dirac delta peaking at the corresponding particle mass.\\

In both cases, the DM density $\rho_{\text{DM}}$ depends on the location of the MC under consideration. For clouds close to the Solar System, a value of $\rho_{\text{DM}} = 0.4 \,\text{GeV}/\text{cm}^3$ is adopted, consistent with recent determinations \cite{Benito_2021}. This value corresponds to the lower edge of the currently allowed range within that work and is thus a conservative choice. In fact, any larger local value of $\rho_{\text{DM}}$ would only strengthen the derived constraints, since it would increase the predicted ionization rate by a factor of $\rho^2$ in the annihilation case and by a factor of $\rho$ in the decay and PBH cases. For MCs located far from the Solar System, the DM distribution within the galaxy is modeled with a Navarro-Frenk-White (NFW) profile \cite{Navarro_1996}, an standard choice for this kind of works. It takes the form:

\begin{equation}
    \rho(r)
  = \frac{\rho_s}{\bigl(r/r_s\bigr)\,(1 + r/r_s)^{2}}
  \label{NFWprofile}
\end{equation}
where $r_s = 20$ kpc \cite{yang2013constraining} is the scale radius of the Milky Way and $\rho_s$ is the characteristic density at $r = r_s$ normalized to have the local value of the DM density at the Solar galactocentric radius. The main targets will be MCs with the lowest possible ionization rate since it will give the most constraining values. The clouds L1551, G28.37+00.07, often referred as DRAGON and G1.4-18+87 were chosen for that task. They present the lowest ionization rate values, but each one of them have their own peculiarities:\\

\emph{L1551} is one of the most promising local clouds for indirect DM searches. It is located towards the Taurus constellation with a size of $\sim 1$ pc and 150 pc away from Earth and has a low ionization rate $\zeta \approx (5 \pm 1.5)\cdot10^{-18} \text{s}^{-1}$ \cite{deboisanger1995ionizationfractiondenseclouds}. As a local cloud, it represents the best target, due to the fact that there is no need of assuming any DM density profile. The measured value can be used directly, with the only source of uncertainty coming from the DM density measurement, yielding robust constraints. Local clouds are, therefore, the best target for this purpose.\\

However, lower values of ionization rate have been found in the so-called Infrared Dark Clouds (IRDC). Even though one should assume a specific density profile, these clouds can highlight the potential of this new observable. As getting closer to the galactic center, the DM density increases. This, combined with the fact that IRDCs exhibit low ionization rate values, the resulting constraints could be improved by up to three orders of magnitude, or even more, depending on the location of the cloud and the chosen profile. One well studied  is the \emph{DRAGON cloud} \cite{Entekhabi_2022}, located at 3.2 kpc from the galactic center with a size $L \sim 13.6$ pc. The previous study of this cloud showed strong inhomogeneities in ionization, with some sub-regions manifesting extremely low values (as low as $10^{-19}$--$10^{-18}\,\mathrm{s}^{-1}$). Since DM-induced ionization should be uniform across the cloud, limits could be derived from the lowest-ionized regions, which provide the most stringent constraints. However, as a conservative choice, in the context of this work it will be taken an upper limit twice the highest value of ionization rate found on that work $\zeta_{DRAGON} = 9.2 \cdot 10^{-19} \text{s}^{-1}$ for setting the constrains.\\

Finally, as an illustrative case it is considered \emph{G1.4-18+87}, a cloud with a size of 8.2 pc located about 400 pc from the Galactic Center and above the galactic plane \cite{Bhoonah_2018}. Being the closest cloud to the center of this work, it offers a significant opportunity to improve current results, as the expected DM density in this region is considerably higher respect to the other clouds. Although earlier works suggested an exceptionally low ionization rate ($\zeta \sim 10^{-19}\,\mathrm{s}^{-1}$) \cite{Bhoonah_2019} , the result is still debated due to uncertainties in the gas temperature \cite{Farrar_2020} . It is therefore adopted $\zeta = 3.8\cdot10^{-19}\,\mathrm{s}^{-1}$ as an optimistic benchmark, assuming an NFW DM profile as in the previous case. This should not be seen as a strict bound, but as an optimistic case in the context of the potential of MCs near the Galactic Center and upper the galactic disk, where CR ionization should be significantly reduced.\\ 

Since a conservative approach is followed, all constraints are derived at the 95\% confidence level. This is implemented by requiring the ionization rate induced by DM emissions to match the 2$\sigma$ upper limit of the measured values. Once the computation is done, the $2\sigma$ upper value from bibliography will be divided by the ionization rate caused by DM obtained from the codes (See Appendix I) and the assumed constants ($\langle\sigma v\rangle, \tau, f_{PBH}$) will be rescaled as follows:

\begin{align}
\begin{cases}
    \dfrac{\zeta^{OBS}}{\zeta^{DM}} = r \longrightarrow \langle\sigma v\rangle_{limit} = \langle\sigma v\rangle_{assumed}\cdot r &\text{(annihilation)}\\[1em]
    \dfrac{\zeta^{OBS}}{\zeta^{DM}} = r \longrightarrow \tau_{limit} = \dfrac{\tau_{asumed}}{r}  &\text{(decay)}\\[1em]
    \dfrac{\zeta^{OBS}}{\zeta^{PBH}} = r \longrightarrow f_{PBH} = r  
    &\text{(PBHs)}
    \end{cases}
\end{align}
Obtaining the final constraints.

\newpage
\section{Results and discussion}
\begin{figure}[h!]
     \begin{subfigure}[b]{0.5\textwidth}
         \centering
         \includegraphics[width=\linewidth]{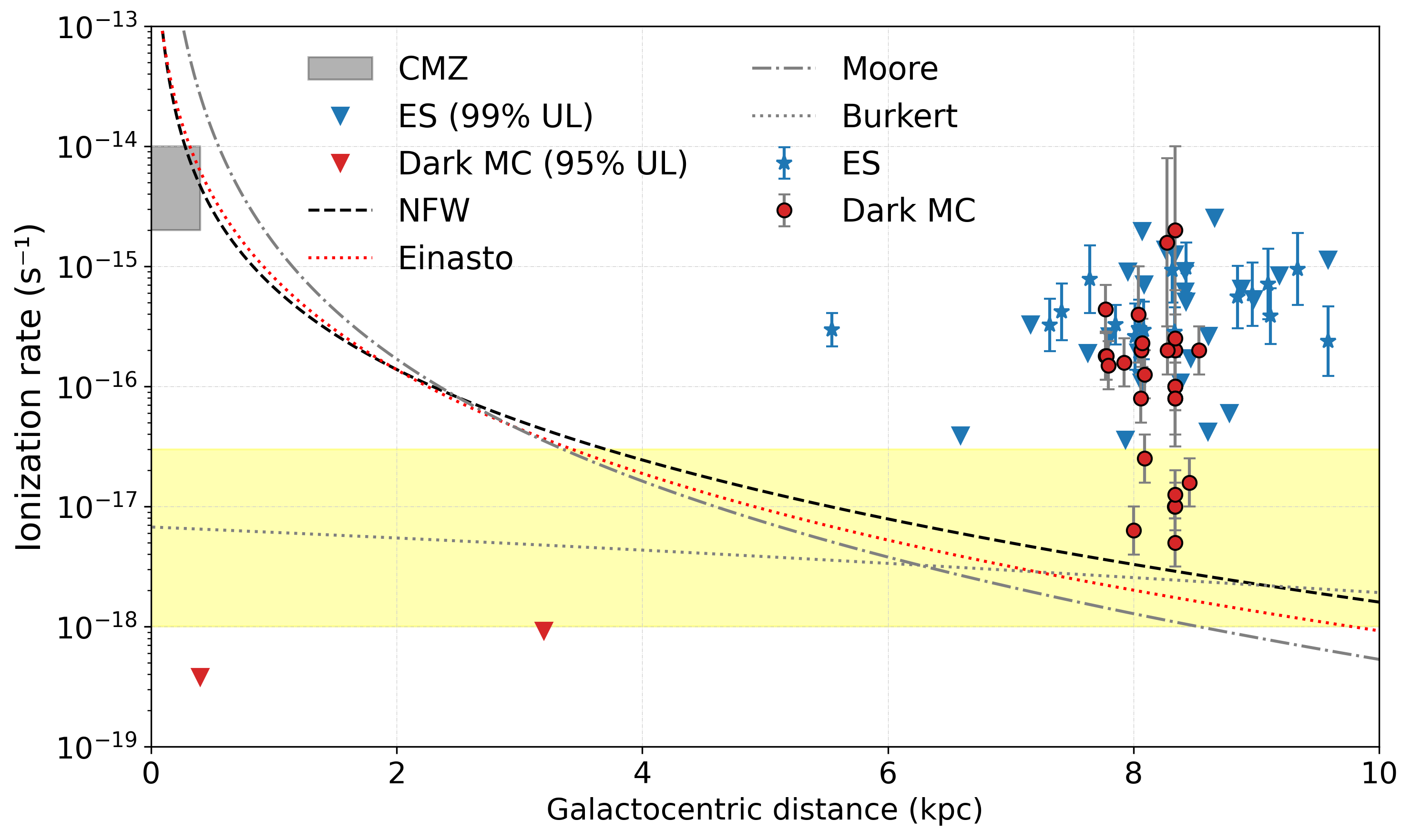}
         \caption{Annihilation fits}
         \label{annihMC}
     \end{subfigure}
     \hfill
     \begin{subfigure}[b]{0.5\textwidth}
         \centering
         \includegraphics[width=\linewidth]{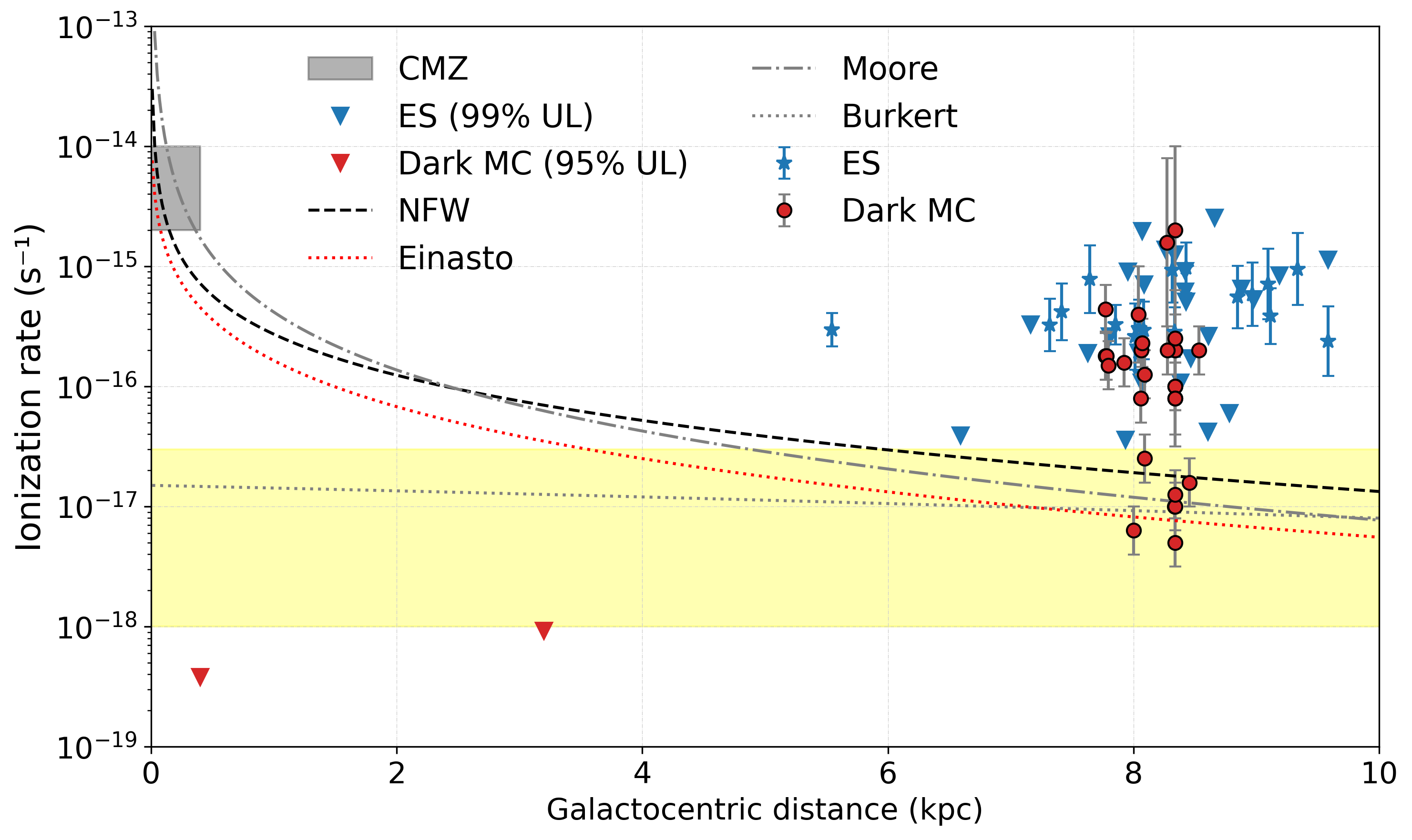}
         \caption{Decay fits}
         \label{decMC}
     \end{subfigure}
        \caption{Ionization rates of various MCs as a function of the distance. The MCs are categorized as follows: \textbf{ES}: MC with embedded stars or star formation (blue circles) \cite{indriolo2012investigating}. \textbf{ES (99\% UL)} are 3$\sigma$ upper limits of ES MC where there is not a best fit (blue triangles) \cite{indriolo2012investigating}. \textbf{Dark MC}: dense MC without star formation within it (red circles) \cite{caselli1998ionization}. \textbf{Dark MC (95\%UL)}: 2$\sigma$ upper limits of dense MC (red triangles). In both plots there are fitted normalized curves of source functions of DM following different DM density profiles. The yellow band accounts for values of ionization rate that could be produced by CRs. The grey rectangle stands for the values of ionization rate within the Central Molecular Zone \cite{Ravikularaman_2025}.}
        \label{MCIRVALUES}
\end{figure}

Assuming that DM annihilation and decay channels into electrons and positrons are allowed, or alternatively considering emissions from evaporating PBHs, it could be expected a correlation between the ionization rates of the MCs and their distance from the center. Since higher presence of DM is expected in the inner parts of the galaxy, this should result on higher values of ionization rates as getting closer to it. 
The most used parameterizations of the DM density distributions are motivated by N-body simulations. These include the NFW profile \cite{Navarro_1996} defined above (Eq. (\ref{NFWprofile})), the Moore profile \cite{moore1998resolving}

\begin{equation}
    \rho_{\text{Moore}}(r) = \frac{\rho_s}{\left(\frac{r}{r_s}\right)^{1.5}\left(1+\frac{r}{r_s}\right)^{1.5}},
\end{equation}
where $\rho_s$ and $r_s$ are the characteristic density and scale radius, respectively, or the Einasto profile  \cite{einasto1965construction}, as representative cases

\begin{equation}
    \rho_{\text{Einasto}}(r) = \rho_s \exp\left\{-\frac{2}{\alpha}\left[\left(\frac{r}{r_s}\right)^{\alpha}-1\right]\right\}
\end{equation}
at which $\rho_s$ is again the density at the scale radius $r_s$, and $\alpha$ is a shape parameter that controls the curvature of the profile.
The first two predict a cusp of density of DM in the center of galaxies that fall off as a power law towards the outer regions, while the Einasto shows a smoother decrease that suits well in the simulations. In contrast, alternative models, such as the Burkert profile, foretell a flatter distribution (cored profile) \cite{burkert1995structure}

\begin{equation}
    \rho_{\text{Burkert}}(r) = \frac{\rho_s}{\left(1+\frac{r}{r_s}\right)\left(1+\left(\frac{r}{r_s}\right)^2\right)}
\end{equation}
where $\rho_s$ represents the central density and $r_s$ the core radius. It is therefore interesting to plot the ionization rate of MCs as a function of the Milky Way galactocentric distance and to compare these measurements with the source functions predicted by different DM density profiles, in order to investigate whether a correlation can be established (Figure (\ref{MCIRVALUES})).\\ 

As can be seen, there is a lack of measurements in the inner regions of the Galaxy. This is mainly because observations at those distances are particularly challenging and carry large uncertainties. Moreover, the detection of absorption bands usually requires the presence of a light background source. For this reason, dark clouds are especially difficult to measure.\\

No clear correlation is observed between all MCs and the source profiles expected from DM annihilation or decay. For MCs with ongoing star formation (ES MCs), very high ionization rates are obtained, as expected, since star formation enhances the ionization and may represent the dominant contribution to the observed values. However, dark MCs also show unexpectedly high ionization rate values, although in some cases these are consistent with the theoretical fits and specially low in the inner regions. In this work, a conservative scenario is assumed in which all the measured ionization is attributed only to DM processes, in order to establish reliable limits. In a more realistic framework, however, interstellar CRs are also expected to contribute significantly to the ionization rate (yellow band). When this contribution is taken into account, the fits show a clearer correlation with theoretical expectations.

\subsection{Constraints on light Dark matter}

\begin{figure}[h!]
     \begin{subfigure}[b]{0.5\textwidth}
         \centering
     \includegraphics[width=\linewidth]{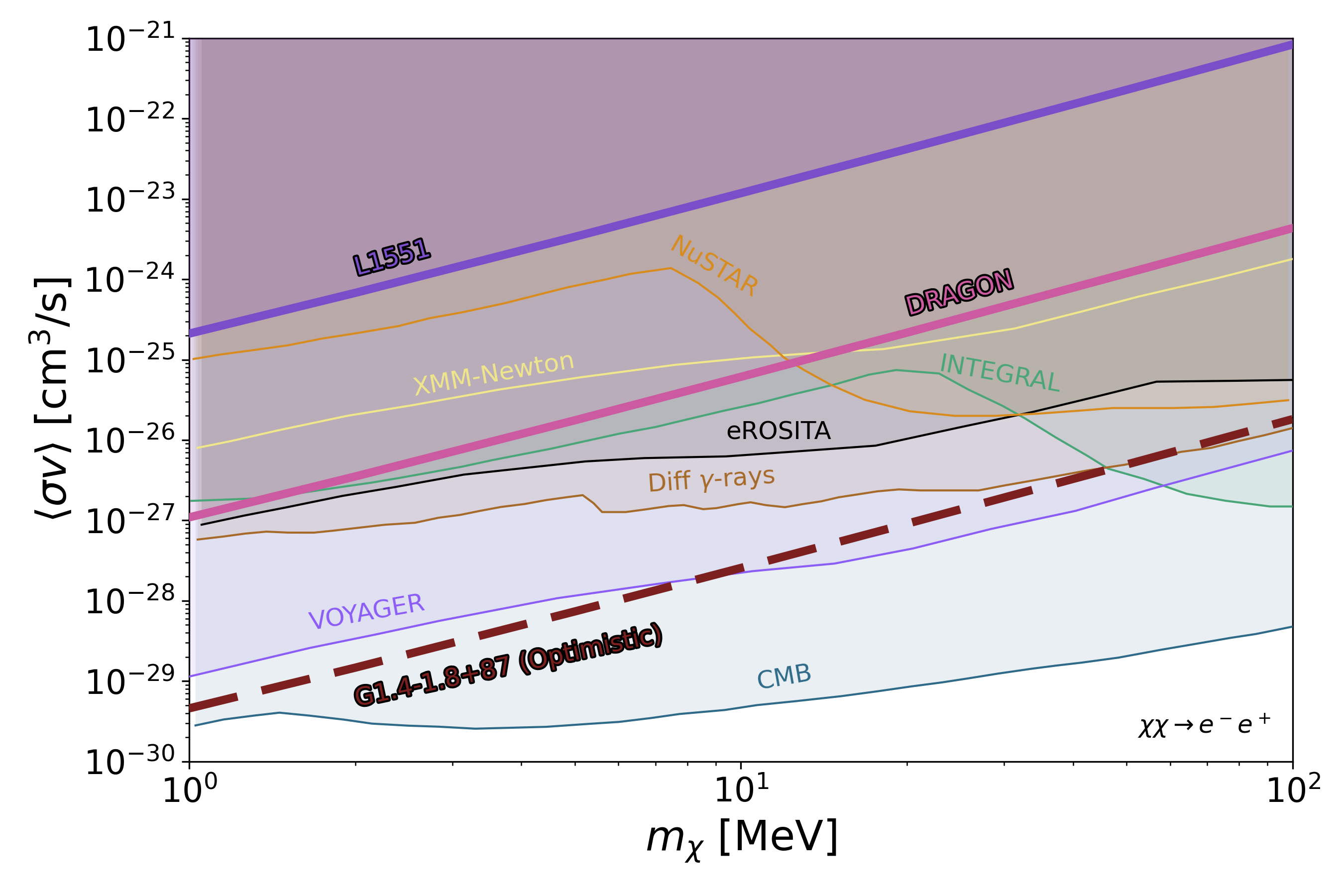}
         \caption{Constraints on annihilating DM}
         \label{annihcons}
     \end{subfigure}
     \hfill
     \begin{subfigure}[b]{0.5\textwidth}
         \centering
         \includegraphics[width=\linewidth]{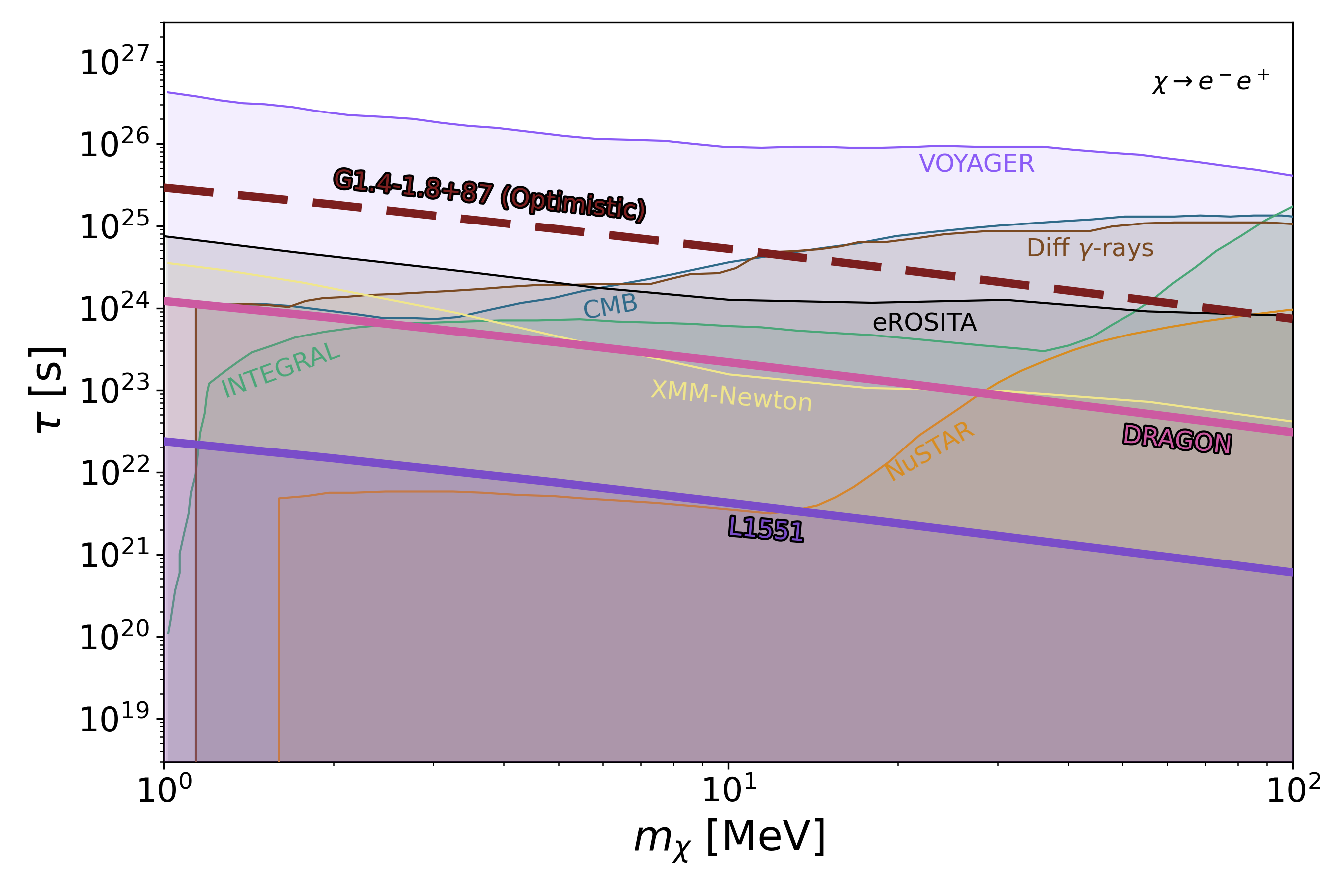}
         \caption{Constraints on decaying DM}
         \label{decaycons}
     \end{subfigure}
    \hfill
        \caption{Comparison of the 95\% confidence levels results derived in this work with existing constraints on the cross section (a) and the decay lifetime (b) as a function of the mass of the DM particle $m_\chi$. Thicker lines with labels highlighted with a black outline are the ones derived in this work while thinner are obtained from other works.}
        \label{contrainsDM}
\end{figure}

As it was said previously, following a very conservative approach, the constrains derived will be at 95\% confidence level and it will be assumed that all the ionization rate is caused only by DM processes, without the CRs contribution. That way, this ionization rate induced by DM (from the simulation) must match a 2$\sigma$ upper value of the observed one.  In figure (\ref{contrainsDM}) it is compared the constraints derived in this work with existing bounds on thermally averaged cross sections and on the lifetime of decaying DM with masses between (1-100) MeV. The results for L1551, G1.4–1.8+87 (Optimistic), and the DRAGON region are shown alongside limits from NuSTAR, INTEGRAL, XMM–Newton, eROSITA, Voyager, CMB, and diffuse $\gamma$–ray observations \cite{Cirelli_2023, Linden_2025, nguyen2024strongconstraintsdarkphoton, balaji2025darkmatterxraysrevised,slatyer2016indirect,lopez2013constraints, essig2013constraining}.\\

\begin{figure}[h!]
        \begin{subfigure}[b]{0.5\textwidth}
         \centering
    \includegraphics[width=\linewidth]{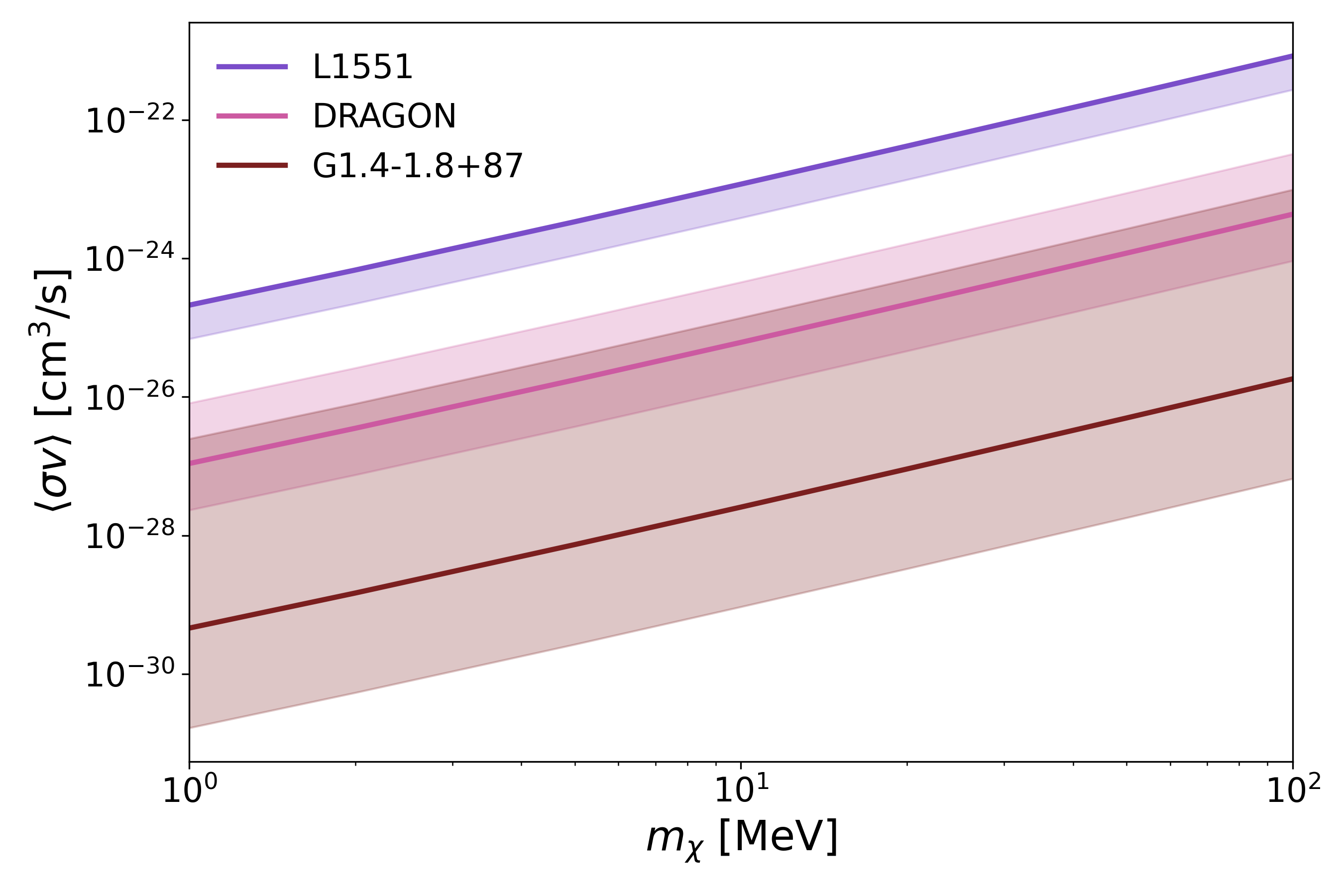}
         \caption{Uncertainties on annihilating DM}
         \label{annihconsuncert}
         \end{subfigure}
    \hfill
    \begin{subfigure}[b]{0.5\textwidth}
         \centering
         \includegraphics[width=\linewidth]{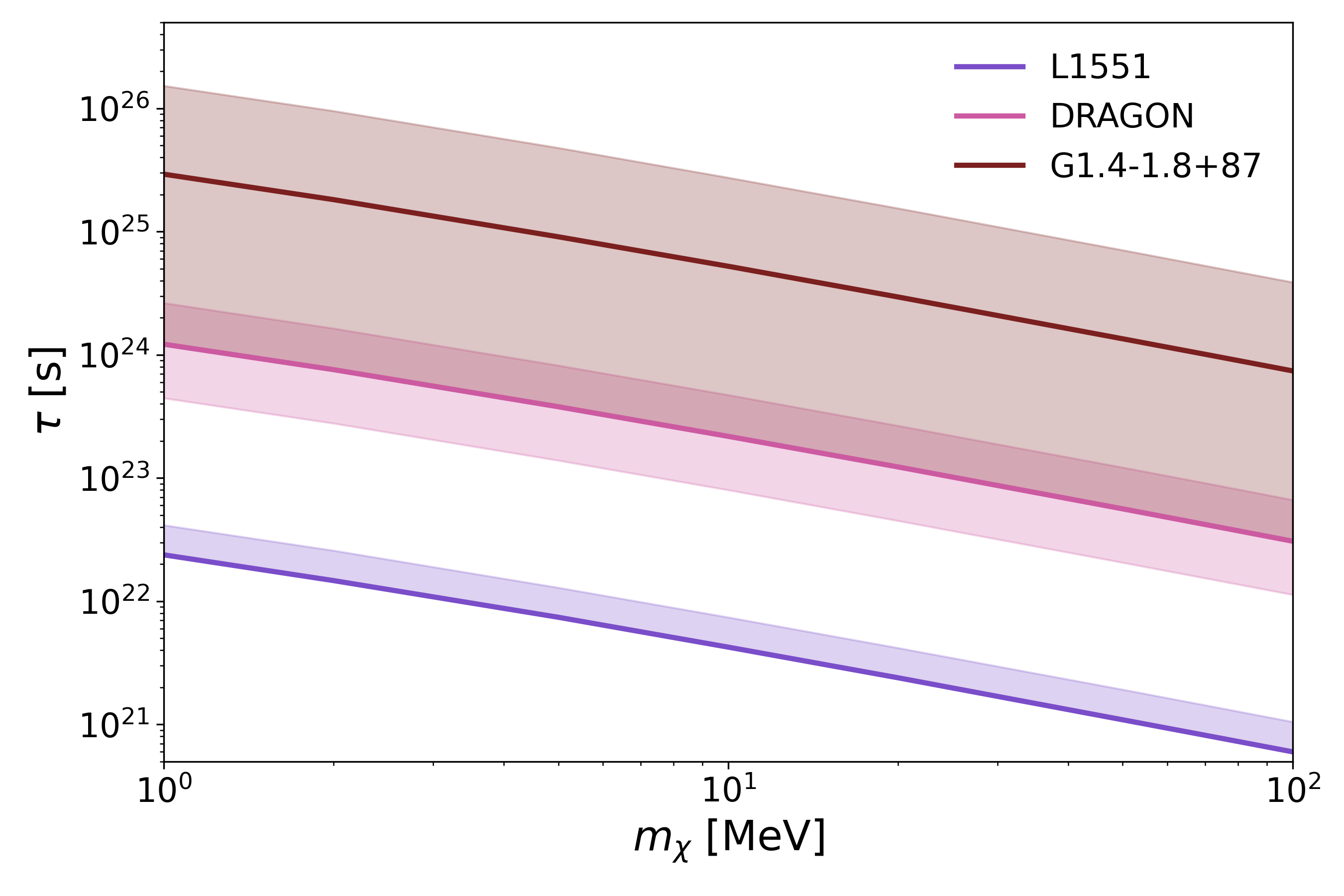}
         \caption{Uncertainties on decaying DM}
         \label{decayconsuncert}
         \end{subfigure}
    \caption{Figures (a) and (b) show the constraints obtained for the annihilating and decaying DM scenarios, respectively. In the L1551 (local) case, the thick line corresponds to a density of $\rho = 0.4  \text{GeV}/\text{cm}^3$, while the uncertainty band is derived by increasing the local DM density up to $\rho = 0.7  \text{GeV}/\text{cm}^3$. For DRAGON and G1.4–1.8+87, the thick line corresponds to the limits obtained under the standard NFW profile assumption, and the associated uncertainties are estimated by replacing the NFW profile with either a Moore with a $\rho = 0.7$ $\text{GeV}/\text{cm}^3$ or a Burkert profile.}
    \label{uncertaintiespart}
\end{figure}

The \textbf{L1551} limits provides the smallest constrains in both cases, yet robust, ruling out cross sections above $\langle \sigma v\rangle \approx 10^ {-25} - 10^{-21}$ $\text{cm}^3/\text{s}$ and decay lifetimes below $\tau \approx 5\cdot10^{22} - 10^{21}$s. It lies significantly below the bests limits like CMB and Voyager but yet surpasses some constrains at lower range of masses (were the approximation made in the code behaves better). For example, in the decaying scenario it gets to improve the NuSTAR up to $m_\chi \approx 20$ MeV and an small region of INTEGRAL, were the sensitivity of the experiment breaks.\\

The \textbf{G1.4–1.8+87 (Optimistic)} scenario yields the strongest sensitivity across the entire mass window considered. Its exclusion band gets to improve existing limits even by three orders of magnitude for $m_\chi \lesssim 10$ MeV, tightening the constraints on DM on annihilation cross sections within $\langle \sigma v\rangle \approx 7\cdot 10^{-29} - 10^{-27}$ $\text{cm}^3/\text{s}$ and decay lifetimes $\tau \approx 5\cdot 10^{25} - 5\cdot 10^{24}$s in this regime. It also keep improving most of the limits in the whole range, but keeps getting worse as the mass increases. Nevertheless, the most reliable results with this method are those below $m_\chi \sim 30$ MeV.  While this estimate should be regarded as optimistic, it nonetheless demonstrates the potential of taking deep observations of MCs for this task.

\textbf{DRAGON} results lie in between the two others. Although it is less restrictive than the optimistic G1.4–1.8+87 case, they remain competitive with established limits from serious experiments. It overlaps probed regions from XMM-Newton, INTEGRAL, diffuse $\gamma$-rays and CMB at low masses, providing also complementary coverage at intermediate masses. It even gets over NuSTAR and XMM.Newton results, showing that this new observable can, in fact, be competitive with other experiments.\\

Overall, the analysis shows that ionization rate of MCs can provide complementary constrains and, in some cases, even stronger that actual X-ray observations and cosmological proves. This highlights the importance of these objects as cosmic laboratories for testing sub-GeV DM processes. The local case is not so competitive, but motivates local searches of clouds with exceptionally small ionization rates in order to improve these results. The DRAGON case starts to compete with other experiments, although it does so at a disadvantage due to the uncertainty introduced by assuming a density profile. Finally, the optimistic case illustrates the values that can be achieved with this method. Nevertheless, as can be observed in Figure (\ref{uncertaintiespart}), the uncertainties of the last two cases are large because measurements of the DM density near the Galactic center are highly unclear, and the assumption of different profiles leads to very different estimations.

\subsection{Constraints on Primordial black holes}

For PBHs, it must be stressed again that it is only considered the contribution from emitted electrons, shown in Figure (\ref{inyrateeandpho}), and not from photons. The reason is because the ionization effect coming from photons becomes relevant only for very small PBH masses, which are already strongly constrained by Hawking evaporation. Therefore, this analysis will be focused on the electron channel as the dominant contribution for this mass window.\\

\begin{figure}[h!]
     \begin{subfigure}[b]{0.5\textwidth}
         \centering
     \includegraphics[width=\linewidth]{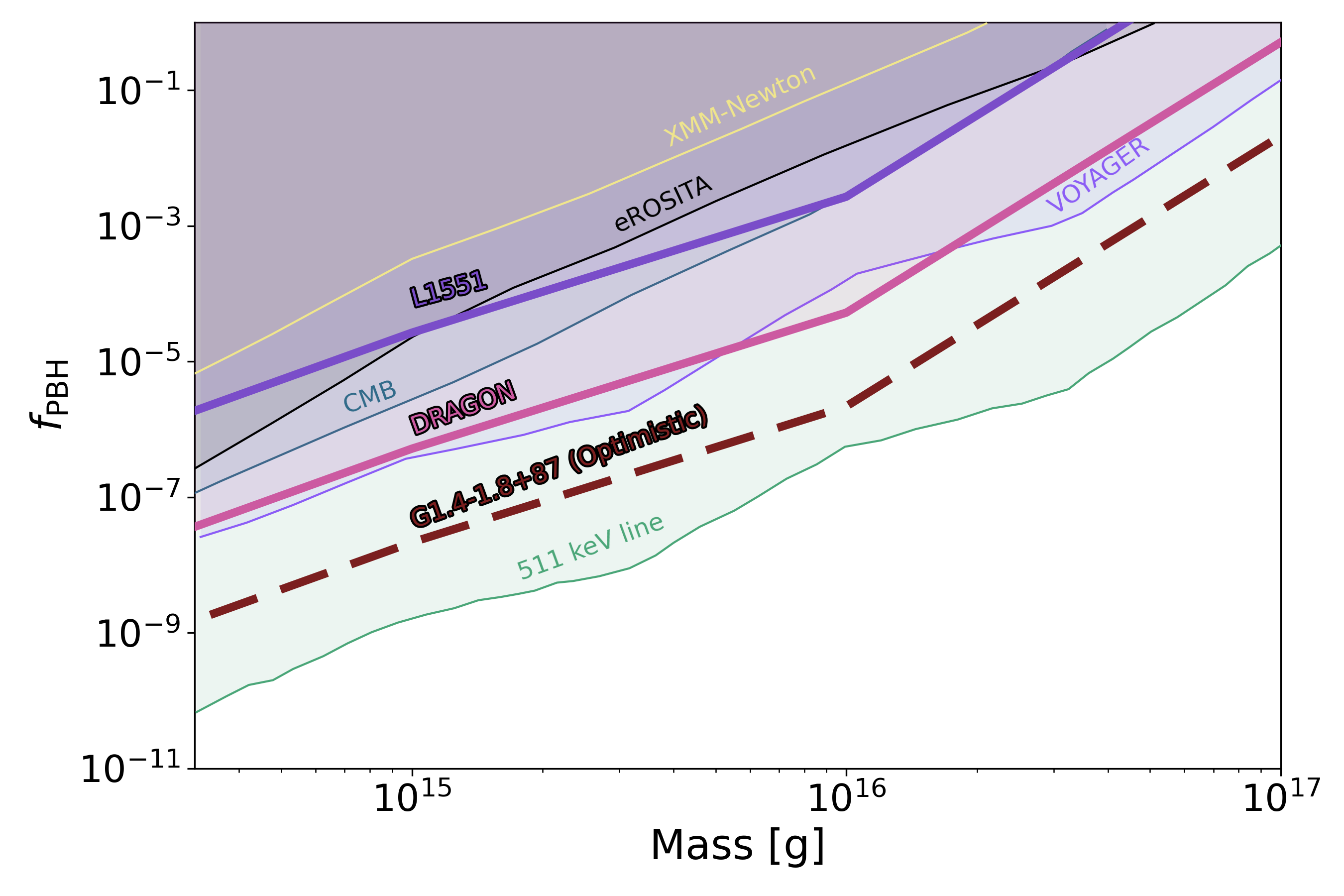}
         \caption{Constraints on PBH abundances}
         \label{pbhplot}
     \end{subfigure}
     \hfill
     \begin{subfigure}[b]{0.51\textwidth}
         \centering
         \includegraphics[width=\linewidth]{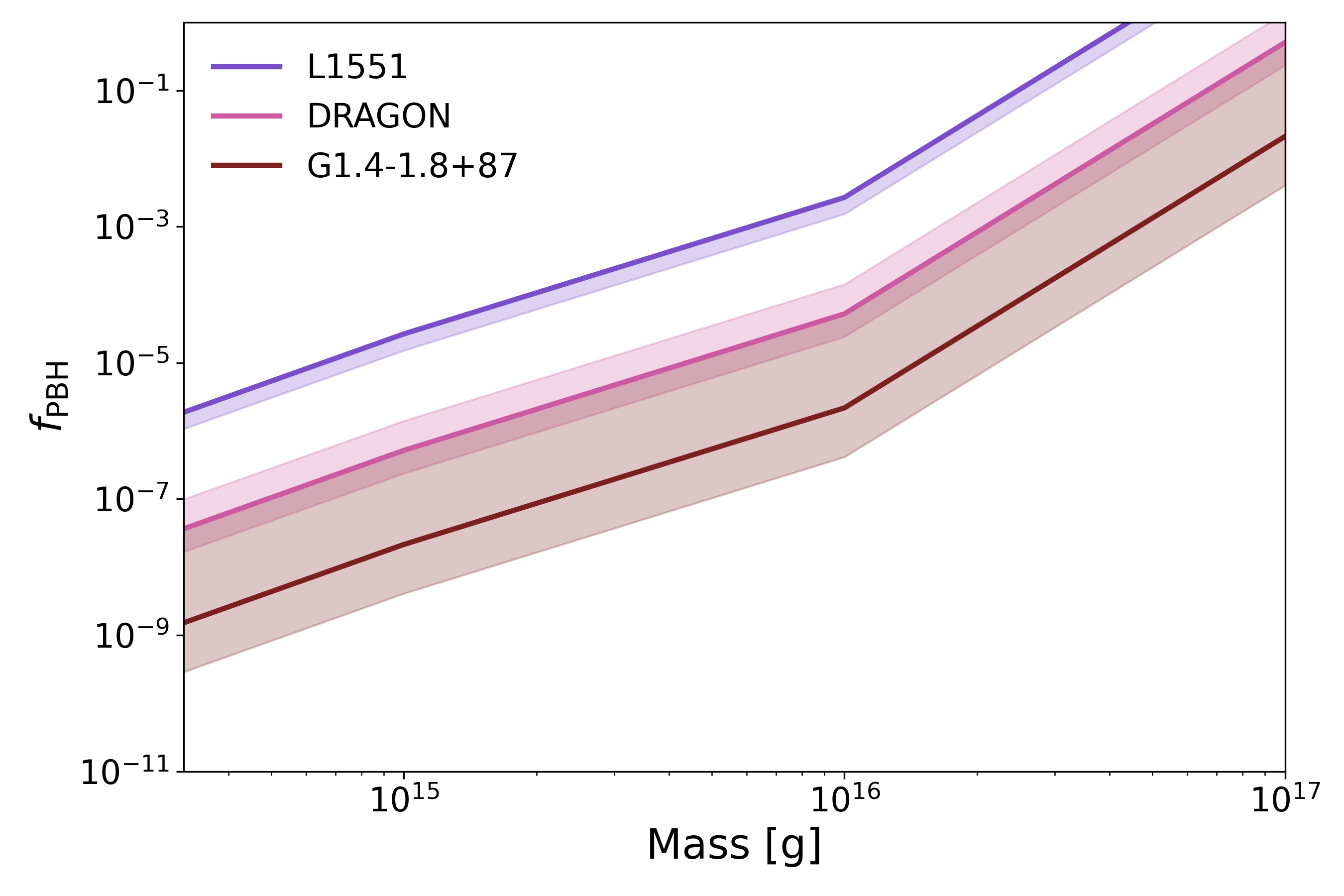}
         \caption{Uncertainties on PBH abundaces}
         \label{pbhunc}
     \end{subfigure}
    \caption{Figure (a) shows the comparison of the 95\% confidence levels results derived in this work with existing constraints on the PBH fraction abundance $f_{PBH}$ depending or their mass. Thicker lines with labels highlighted with a black outline are the ones derived in this work while thinner are obtained from other works \cite{luque2024refininggalacticprimordialblack, acharya2020cmb, chluba2020thermalization, balaji2025darkmatterxraysrevised}. Figure (b) shows the constraints obtained for PBH abundances. In the L1551 (local) case, the thick line corresponds to a density of $\rho = 0.4 \text{GeV}/\text{cm}^3$, while the uncertainty band is derived by increasing the local DM density up to $\rho = 0.7  \text{GeV}/\text{cm}^3$. For DRAGON and G1.4–1.8+87, the thick line corresponds to the limits obtained under the standard NFW profile assumption, and the associated uncertainties are estimated by replacing the NFW profile with either a Moore with a $\rho = 0.7$ $\text{GeV}/\text{cm}^3$ or a Burkert profile.}
    \label{pbh constrains}
\end{figure}

Figure (\ref{pbhplot}) shows the comparison between the constrains obtained with the MCs and existing bounds on PBHs abundances in the asteroid mass range. The vertical axis shows the fraction of DM in the form of PBHs, $f_{PBH}$, while the horizontal axis indicates the PBH mass in grams. In this context, lower values of $f_{PBH}$ correspond to stronger (i.e. more restrictive) constraints, since they imply that only a very small fraction of DM can be composed of PBHs of a given mass.\\

Among the targets proposed, \textbf{L1551} provide limits that are pretty competitive, specially in the intermediate mass range around $ 10^{16}$g, improving the results of XMM-Newton, eROSITA, and even matching CMB constrains beyond that mass. However, it still lies above three orders of magnitude away from the strongest constrains, like the 511 keV line.\\

\textbf{DRAGON} constrains get to improve most of the limits on the whole range. It overlaps with VOYAGER limits, providing a complementary coverage to it. Although it keeps away from the most stringent limit in the graph, it shows that this method can compete on setting possible PBH contributions to DM.\\

The \textbf{G1.4-1.8+87 (Optimistic)} case again yields the strongest result of all three over the entire mass window. It rules out contributions of PBHs above $10^{-9}$ on the low mass regime and let bigger contributions as the mass of the black hole increases. Despite being optimistic, such target illustrates the strength of the method, allowing it to compete with the strongest constraints, nearly matching the 511 keV line limit.\\

Again, the chosen MCs seem also as valuable targets for setting the PBHs contributions to DM. The electrons emitted through Hawking's radiation would, in fact, produce this astrophysical signature. The resulting limits can cover part of the asteroid mass PBHs range, with the optimistic scenario illustrating how, under favorable assumptions, this method can reach sensitivities close to the strongest existing bounds. The use of MCs as targets demonstrates a powerful approach to significantly reduce the available parameter space for PBHs as DM candidates.

\newpage
\section{Conclusions}
The work developed in this thesis has explored the potential of MCs as novel laboratories for constraining sub-GeV DM models as well as PBHs. Focusing on the ionization rates as the observable, this study has proven that MCs, often studied as sites for star formation, can also provide strong constrains for indirect DM searches.\\

The methodology adopted followed a very conservative approach. It was assumed that all the ionization rate was produced only by DM-induced electrons and positrons. That way, it is ensured that the bounds derived are robust and not artificially improved by uncertain astrophysical backgrounds. Even though this assumption excludes contributions from the CRs population, it guarantees that the results are reliable despite the particular location of the cloud and/or the cosmic ray population around. Also, the DM–induced ionization rate was required to match the $2\sigma$ upper value of the observed ionization rate, corresponding to a 95\% confidence level. This choice ensures consistency with the convention used in the literature when reporting upper limits. \\

For light DM in the MeV regime, the analysis showed that MCs can set constraints that are competitive with, and in some scenarios even superior to, those obtained from X-ray experiments or cosmological probes. In particular, regions such as the DRAGON cloud and the optimistic chosen case of G1.4–1.8+87 provide exclusion bounds on annihilation cross sections and decay lifetimes that compete with existing limits in the low-mass window, where the approximation of the transport equation behaves better. This highlights the extraordinary potential of clouds located near the galactic center. Even in the local case of L1551, the bounds obtained, although less restrictive, remain solid and independent of halo profile assumptions. The evidence proves that nearby MCs with exceptionally low ionization values are prime targets for future indirect searches.\\

In the PBH scenario, the results are also promising. By considering the electrons coming from Hawking's radiation as the dominant component for ionizing the gas, the present work has shown that the ionization rate of MCs can place competitive constrains on the PBH fraction of DM in the asteroid mass range. As it was the case in the particle candidate, the local cloud gives complementary but less competitive limits. However, inner galaxy candidates like DRAGON or G1.4–1.8+87 begin to be competitive. Specially, in the optimistic case, surpasses most of the limits, even reaching at a certain range one of the most stringent bound as the 511 keV line is. These findings suggest that MCs can act as observables to constrain the remaining parameter space for PBHs as viable DM candidates.\\

Despite the numerical results, this thesis establishes an important conceptual conclusion: MCs are, indeed, a new astrophysical observable that can contribute to the indirect DM detection efforts. While CRs have long been considered as the dominant component for ionization, the persistent anomalies and unexplained excesses observed in both diffuse and dense clouds motivate exotic contributors. DM, whether in the form of light particles or evaporating PBHs, can offer a natural explanation for these excesses. Hence, this analysis can establish a possible framework to quantify these contributions.\\

This work opens a broad range of future researches. One of the most important is the extension of observational experiments aimed to measure ionization rates of MCs across the Milky Way. In particular, in the low sampled inner parts of the Galaxy, but also potentially low ionized local clouds. Improving the precision of these measurements, will directly make better the robustness of the constrains. On the theoretical aspects, the incorporation of more realistic treatments of cloud structure, diffusion processes, turbulences or even multi-channel DM interactions would test the reliability of the assumptions made along the work. In addition, a combination of DM-injected electrons and positrons with a cosmic ray population as ionizing components could lead to a huge improvement in the derived constrains, but perhaps less robust, since it would be added an extra uncertainty coming from the measured CRs.\\

Moreover, the approximation implemented in the transport equation leaves itself space for a significant improvement. As it was explained in the method section, the diffusion term within the cloud was neglected in favor of conservative approximation where energy losses dominate. This assumption is justified for compact and dense clouds and with low-energy electrons and positrons. More realistic simulations could incorporate spatial diffusion and turbulence. In the context of PBHs, decreasing mass is expected to yield more energetic electrons and positrons, at which point the approximation begins to break down. At higher energies, electrons and positrons are more likely to escape from the clouds therefore, it cannot be assumed that they deposit all of their energy within them. Refining the code to include these effects, together with a better treatment of the cloud geometry and density profiles (not just homogeneous gas density) would provide more accurate predictions thus stronger constrains.\\

Finally,  it must be highlighted that this work points toward several promising future researches. First, the methodology developed here can be extended beyond the scenario of annihilating and decaying DM and evaporating PBHs to other candidates such as axion like particles or more complex hidden sectors. These other DM candidates could also inject energy into MCs through novel channels, and the same framework could be adapted to such scenarios. That way, the strategy established in this thesis not only demonstrates the present utility of MCs as DM detectors, but also gives a possible way for enhancing their role in the future. By considering more candidates and improving the modeling of astrophysical environments, MCs could become a valuable tool for probing the dark sector.

\newpage
\addcontentsline{toc}{section}{References} 


\bibliographystyle{IEEEtran} 
\bibliography{refsTFM} 

\newpage
\addcontentsline{toc}{section}{List of figures}
\listoffigures

\newpage
\renewcommand{\thesection}{A\arabic{section}}
\renewcommand{\thesubsection}{A\arabic{subsection}}
\section*{Appendix I}\addcontentsline{toc}{section}{Appendix I}
\subsection*{DM particles induced ionization rate}
Below it is written the code for computing the induced ionization rate from DM particles processes.

\begin{Verbatim}[breaklines=true]
    import matplotlib.pyplot as plt
from astropy.io import fits as pyfits
import numpy as np
from matplotlib import rcParams
import os
# from matplotlib.colors import LogNorm
import pylab
from scipy import integrate
import sys
from scipy.optimize import curve_fit

# Useful variables
#erg_to_GeV = 624.151  # GeV/erg
kpc_to_cm = 3.086e21  # cm/kpc
AU_to_cm = 1.5e13  # cm/AU Astronomical units
yr_to_sec = 3.154e7  # s/yr
m_e = 0.511 * 1e-3  # GeV
m_p = 0.938  # GeV
c_light = 29979245800  # cm/s
hbar = 4.135667696 * 10 ** (-24) / (2 * np.pi)  # GeV*s
alpha = 1 / 137.03
a0 = c_light * hbar / (m_e * alpha)  # cm #5.291772e-9 # cm


def Gamma_e(Ekin_e):  ## must be in MeV
    E_e = (Ekin_e + (m_e * 1e3))
    return E_e / (m_e * 1e3)


def Beta_e(Ekin_e):
    gamma = (Ekin_e + (m_e * 1e3)) / (m_e * 1e3)
    return np.sqrt(1 - (1 / (gamma ** 2)))


I_H = 13.6 * 1e-9  # eV --> GeV
I_He = 24.59 * 1e-9  # eV --> GeV


def DE_dt(Evec_final, myn_H):  # DRAGON
    Mygamma = Gamma_e(Evec_final)
    Mybeta = Beta_e(Evec_final)
    Ion_H = 7.64e-18 * myn_H / Mybeta ** 2 * (np.log(Mygamma ** 4 * m_e ** 2 / (2 * I_H ** 2 * (1 + Mygamma))) - (
                2 / Mygamma - 1 / Mygamma ** 2) * np.log(2) + 1 / Mygamma ** 2 + 1 / 8 * (
                                                          1 - (1 / Mygamma)) ** 2) * 1000  # MeV/s
    Ion_He = 2 * 7.64e-18 * 0.11 * myn_H / Mybeta ** 2 * (
                np.log(Mygamma ** 4 * m_e ** 2 / (2 * I_He ** 2 * (1 + Mygamma))) - (
                    2 / Mygamma - 1 / Mygamma ** 2) * np.log(2) + 1 / Mygamma ** 2 + 1 / 8 * (
                            1 - (1 / Mygamma)) ** 2) * 1000  # MeV/s
    return (Ion_H + Ion_He)


def sigma_eH2(E):
    N = 2
    I_H = 13.598 * 1e-6  # MeV
    I_H2 = 15.603 * 1e-6  # MeV
    t = E / I_H2
    sigmas = 4 * np.pi * (a0 ** 2) * N * (I_H / I_H2) ** 2 * F(t) * G(t)
    sigmas[E < I_H2] = 0.
    return sigmas  # return 4*np.pi*(a0**2)*N*(I_H/I_H2)**2*F(t)*G(t)  # cm^2


def F(t_):
    n = 2.4
    return (1 - t_ ** (1 - n)) / (n - 1) - (2 / (1 + t_)) ** (n / 2) * (1 - t_ ** (1 - (n / 2))) / (n - 2)


def G(t_):
    A1 = 0.74
    A2 = 0.87
    A3 = -0.6
    return (A1 * np.log(t_) + A2 + A3 / t_) / t_

    
Rsch = (2 * 6.67 * 1e-11 * 4.3e6 * 1.989 * 1e30) / (3 * 1e8) ** 2 * 3.241 * 1e-20 * 1e3  # in pc


def SourceDM(gamma_, sv, mDM, E_):  # mDM and E must be in MeV

    Av_DM_Density = 44.7 * 1e3  ## MeV cm^-3
    Inject_val = ((Av_DM_Density/mDM)*1/tau)
    # ((Av_DM_Density/mDM)*1/tau) decay
    #(Av_DM_Density / mDM) ** 2 * 0.5 * sv annih
    # (Av_DM_Density/mDM)**2 * 0.5* sv #* 1/mDM  #dN_dE(E_, mDM) #cambiar para decay definir 1/tau
    DMi = np.argmin(np.abs(mDM-(m_e*1e3)*0.5 - E_))  # entre dos para decay (mDM-(m_e*1e3)*0.5
    #(mDM - (m_e * 1e3) - E_) annih
    mySource = np.zeros(len(E_))
    mySource[DMi] = Inject_val
    return mySource  ## parts cm^-3 s^-1 MeV^-1 --- Source term as function of kinetic energy

    tau = 1e25
sigmav = 1e-30  # cm^3/s  Reference for 10 MeV particle in a Moore profile
M_DM = [1, 2, 5, 10, 20, 50, 75, 100][:]  # MeV
Myslope = 1  # .5
Ref_IoR = 2e-14  # s^-1

E_Sec_Ion = np.array(
    [0.000900778852669541, 0.002106157811188177, 0.00464884915543887, 0.011813752791321823, 0.04196477012610897,
     0.12913093719765525, 0.8191389474756182, 6.4671830191989415, 43.070688898157584, 198.93681085703028,
     387.49550292067363, 1688.837558470262, 4590.058089821797, 11797.936670291001, 41203.42646454589])  # MeV
Sec_Ion = np.array(
    [0.4112, 0.4975999999999999, 0.5712000000000003, 0.6544, 0.7376, 0.7919999999999998, 0.8303999999999999,
     0.8367999999999999, 0.8239999999999998, 0.7888000000000001, 0.7663999999999999, 0.7152000000000002,
     0.6799999999999999, 0.6447999999999996, 0.6063999999999997])  # adimenssional

# exit()
nH_CMZ = 100  # cm^-3
plt.figure(figsize=(8, 6))
colors = plt.cm.plasma(np.linspace(0, 1, len(M_DM)))
for m_DM in M_DM:
    Evec = np.logspace(-6, np.log10(5 * m_DM), 100)  # MeV
    dphi_dE = np.sum(SourceDM(Myslope, sigmav, m_DM, Evec)) / DE_dt(Evec,
                                                                    nH_CMZ)  # cm^-3 s^-1 MeV^-1 * MeV  * s/MeV ->  cm^-3/MeV   #np.trapz(SourceDM(Size, Myslope, sigmav, m_DM, Evec), Evec)/ DE_dt(Evec, nH_CMZ)   # cm^-3 s^-1 MeV^-1 * MeV  * s/MeV ->  cm^-3/MeV
    J = dphi_dE * Beta_e(Evec) * c_light / (4 * np.pi)  ## cm^-2 s^-1 MeV^-1 sr^-1
    # print(np.interp(0.1, Evec, J), sigma_eH2(np.array([0.1])))
    # exit()
    Ioniz_rate = 2 * 4 * np.pi * np.trapz(J * sigma_eH2(Evec) * (1. + np.interp(Evec, E_Sec_Ion, Sec_Ion)),
                                          Evec)  ## 2 to account for both, e- and e+
    # print(m_DM, 'MeV  ', Ioniz_rate/Ref_IoR)
    print(Ioniz_rate)  # /Ref_IoR)
\end{Verbatim}

\subsection*{DM as PBHs induced ionization rate}
The same as before, but for Hawking radiation instead of annihilation and decay sources:

\begin{Verbatim}[breaklines=True]
    import matplotlib.pyplot as plt
from astropy.io import fits as pyfits
import numpy as np
from matplotlib import rcParams
import os
# from matplotlib.colors import LogNorm
import pylab
from scipy import integrate
import sys
from scipy.optimize import curve_fit

# Useful variables
#erg_to_GeV = 624.151  # GeV/erg
kpc_to_cm = 3.086e21  # cm/kpc
AU_to_cm = 1.5e13  # cm/AU Astronomical units
yr_to_sec = 3.154e7  # s/yr
m_e = 0.511 * 1e-3  # GeV
m_p = 0.938  # GeV
c_light = 29979245800  # cm/s
hbar = 4.135667696 * 10 ** (-24) / (2 * np.pi)  # GeV*s
alpha = 1 / 137.03
a0 = c_light * hbar / (m_e * alpha)  # cm #5.291772e-9 # cm


def Gamma_e(Ekin_e):  ## must be in MeV
    E_e = (Ekin_e + (m_e * 1e3))
    return E_e / (m_e * 1e3)


def Beta_e(Ekin_e):
    gamma = (Ekin_e + (m_e * 1e3)) / (m_e * 1e3)
    return np.sqrt(1 - (1 / (gamma ** 2)))


I_H = 13.6 * 1e-9  # eV --> GeV
I_He = 24.59 * 1e-9  # eV --> GeV


def sigma_eH2(E):
    N = 2
    I_H = 13.598 * 1e-6  # MeV
    I_H2 = 15.603 * 1e-6  # MeV
    t = E / I_H2
    sigmas = 4 * np.pi * (a0 ** 2) * N * (I_H / I_H2) ** 2 * F(t) * G(t)
    sigmas[E < I_H2] = 0.
    return sigmas  # return 4*np.pi*(a0**2)*N*(I_H/I_H2)**2*F(t)*G(t)  # cm^2


def F(t_):
    n = 2.4
    return (1 - t_ ** (1 - n)) / (n - 1) - (2 / (1 + t_)) ** (n / 2) * (1 - t_ ** (1 - (n / 2))) / (n - 2)


def G(t_):
    A1 = 0.74
    A2 = 0.87
    A3 = -0.6
    return (A1 * np.log(t_) + A2 + A3 / t_) / t_


# DM profiles
DM_local_rho =  0.64*1e3    #0.4 * 1e3  # GeV/cm^3  --> MeV/cm^3
Scale_R = 20 * 1e3  # pc
Rsch = (2 * 6.67 * 1e-11 * 4.3e6 * 1.989 * 1e30) / (3 * 1e8) ** 2 * 3.241 * 1e-20 * 1e3  # in pc


# lee total_electron_spectra.txt delimitado por comas
Fluxes = np.loadtxt('total_electron_spectra.txt', delimiter=',', skiprows=1, usecols=(1,2,3,4))
E_PBH   = np.loadtxt('total_electron_spectra.txt', delimiter=',', skiprows=1, usecols=(0,)) * 1e3  # → MeV

# lista de masas PBH en gramos y conversión a MeV
M_PBH_g       = np.array([1e14, 1e15, 1e16, 1e17])  # g
MPBHg_to_MeV  = 5.617e26                            # MeV por gramo
M_PBH_list_MeV = M_PBH_g * MPBHg_to_MeV             # MeV

def PBH_rate(mass_MeV, E):
    idx = np.argmin(np.abs(M_PBH_list_MeV - mass_MeV))
    return np.interp(E, E_PBH, Fluxes.T[idx])

def SourcePBH(fraction, mass_MeV, E):
    # número de PBH por cm³
    n_PBH = fraction * DM_local_rho / mass_MeV
    # tasa inyectada [cm^-3 s^-1 MeV^-1]
    return n_PBH * PBH_rate(mass_MeV, E)

# ——————————————————————————————————————————
# Parámetros de ionización secundaria
# ——————————————————————————————————————————
E_Sec_Ion = np.array([0.0009008,0.0021062,0.0046488,0.0118138,0.0419648,
                      0.129131,0.819139,6.467183,43.07069,198.9368,
                      387.4955,1688.8376,4590.0581,11797.9367,41203.42646])
Sec_Ion   = np.array([0.4112,0.4976,0.5712,0.6544,0.7376,0.7920,0.8304,
                     0.8368,0.8240,0.7888,0.7664,0.7152,0.6800,0.6448,0.6064])

# ——————————————————————————————————————————
# Bucle principal: calcula io vs masa (g)
# ——————————————————————————————————————————
nH_CMZ = 100  # cm^-3
Evec   = np.logspace(0, 3, 100)  # de 1 a 1000 MeV

zetas = []
for m_g, m_MeV in zip(M_PBH_g, M_PBH_list_MeV):
    Q    = SourcePBH(1.0, m_MeV, Evec)
    numer_total = np.trapz(Q, Evec)
    dphi_dE = numer_total / DE_dt(Evec, nH_CMZ)
    J    = dphi_dE * Beta_e(Evec) * c_light / (4*np.pi)
    zeta = 4*np.pi * np.trapz(
               J * sigma_eH2(Evec) * (1 + np.interp(Evec, E_Sec_Ion, Sec_Ion)),
               Evec
           )
    zetas.append(zeta)
    print(f"PBH mass = {m_g:.0e} g →  = {zeta:.3e} cm⁻³ s⁻¹")

\end{Verbatim}
\end{document}